\documentclass[AMA,STIX1COL]{WileyNJD-v2}
\usepackage{amssymb}
\usepackage{amsmath}
\usepackage{etoolbox}
\usepackage{amssymb}
\usepackage{graphicx}
\usepackage{amssymb,amsmath,latexsym}
\usepackage{boondox-cal}
\articletype{RESEARCH ARTICLE}%

\begin{document}

\title{Nash Social Distancing Games with Equity Constraints: How Inequality Aversion Affects the Spread of Epidemics\protect\thanks{I.~ Kordonis e-mail: jkordonis1920@yahoo.com,  A.-R. ~Lagos e-mail: lagosth@mail.ntua.gr,  G.P. ~Papavassilopoulos e-mail: yorgos@netmode.ntua.gr}}

\author[1]{Ioannis Kordonis*}

\author[1]{Athanasios-Rafail  Lagos}

\author[1,2]{George P. ~Papavassilopoulos}

\authormark{I. Kordonis, A.-R. Lagos, G.P. Papavassilopoulos }

\address[1]{\orgdiv{School of Electrical and Computer Engineering}, \orgname{National Technical University of Athens}, \orgaddress{\state{9 Iroon Polytechniou str., Athens, Postal Code 157 80}, \country{Greece}}}

\address[2]{\orgdiv{Department of Electrical Engineering--Systems}, \orgname{University of Southern California
 }, \orgaddress{\state{3740 McClintock Ave, Los Angeles, CA 90089}, \country{United States}}}

\corres{* \email{jkordonis1920@yahoo.com}}

\abstract[Summary]{In this paper, we present a game-theoretic model describing the voluntary social distancing during the spread of an epidemic. The payoffs of the agents depend on the social distancing they practice and on the probability of getting infected. We consider two types of agents, the non-vulnerable agents who have a small cost if they get infected, and the vulnerable agents who have a higher cost.   For the modeling of the epidemic outbreak, we consider  a variant of the SIR (Susceptible-Infected-Removed) model involving populations of susceptible, infected and removed persons of vulnerable and non-vulnerable types. The Nash equilibria of this social distancing game are studied. The main contribution of this work is the analysis of the case where the players, desiring to achieve a low social inequality, pose a bound on the variance of the payoffs. In this case, we introduce and characterize a notion of Generalized Nash Equilibrium (GNE) for games with a continuum of players. Through numerical studies, we show 	that inequality constraints result in a slower spread of the epidemic and an improved cost for the vulnerable players. Furthermore, it is possible that inequality constraints are beneficial for non-vulnerable players as well.}


\maketitle


 \section{Introduction}
 \sloppy 
 Epidemics harass humanity for centuries, and people investigate several strategies to contain them. The development of medicines and vaccines and the evolution of healthcare systems with specialized personnel and equipped hospitals have significantly affected the spread of many epidemics and have even eliminated some contagious diseases.  However, during the current COVID-19 pandemic, due to the lack or scarcity of appropriate medicines and vaccines, Non-Pharmaceutical Interventions (primarily social distancing) have been among the most effective strategies to reduce the disease spread. Due to the slow roll-out of the vaccines, their uneven distribution, the emergence of  SARS-CoV-2  variants, age limitations, and people's resistance to vaccination, social distancing is likely to remain significant in a large part of the globe for the near future. 

Epidemiological models are essential in designing measures and strategies to control epidemics\footnote{For example,  Imperial College London's report\cite{ferguson2020report}                                                                                      profoundly influenced UK's response to COVID-19 epidemic.  }. In the last century, epidemiologists have made significant progress in the mathematical modeling of the spread of epidemics. From the seminal works of  Kermack and McKendrick\cite{Kermack} and  Ross\cite{Ross}, a prevalent approach in the mathematical modeling of epidemics is compartmental models. These models consider that each agent belongs in some compartment according to her infection state (e.g., Susceptible-Infected-Recovered) and study the evolution of each compartment's population. The literature on these models is extensive, so for a summary, we refer  to Chapter $2$ of Allen et al.(2008) \cite{allen2008mathematical}. There are also other elegant approaches to epidemics modeling, such as the ones that take into consideration the heterogeneous networked structure of human interconnections\cite{Pastor-Satorras}. However, the compartmental models remain a well-studied and fruitful approach, widely used in real-life applications.

The development of epidemiological models is a valuable tool in designing protective measures against the spread of an epidemic. Still, these measures will be adopted by agents who act in a self-interested manner,  at least to some extent. Thus, game theory is an appropriate complementary mathematical tool to be used in this field. Indeed, many game-theoretic models have been developed to study voluntary vaccination\cite{Zhang1,Chang,Bauch1,Bauch2,Reluga1,Reluga2,Zhang2,Fine-Clarkson} and   behavioral changes of the agents \cite{Kremer,Vardavas,Del_Valle,Chen2,Funk-Review,Funk1,Chen1,d'Onofrio}, such as social distancing, use of face masks, and better hygiene practice. Another closely related stream of research is the study of the adoption of decentralized protection strategies in engineered and social networks\cite{theodorakopoulos2012selfish,trajanovski2015decentralized,hota2019game,huang2019differential}.   
Recently, with the emergence of the COVID-19 pandemic, there is a renewed interest in modeling individual behaviors. Related tools include dynamic game analysis of social distancing\cite{toxvaerd2020equilibrium}, evolutionary game theory\cite{karlsson2020decisions,amaral2020epidemiological,ye2020modelling,kabir2020evolutionary} and  network game models \cite{lagos2020games}.

The majority of game-theoretic models are based on the assumption that the rational behavior for an agent is to maximize selfishly her own payoff ignoring  the social impacts (externalities) of this choice. This can lead to `free-riding' phenomena in vaccination games\cite{Bauch2,Zhang2} or in disobedience of social distancing rules in social distancing games, which can both result in a higher prevalence of the spread of the epidemic\cite{Bauch2,van_Boven} and to harmful consequences for the vulnerable members of the society. The phenomenon that the Nash equilibrium strategies result to a social welfare less than the optimal one is well known in game theory community as the \textit{Tragedy of the Commons}\cite{hardin1968tragedy}. There are some notable exceptions\cite{alfaro2020social,Brown}  analyzing epidemic games involving  altruistic individuals.
However, even in the cases considering that the agents prefer the strategies that maximize a social welfare function, there may still exist significant inequalities among their payoffs.

Epidemics may create vastly unequal outcomes in terms of health risks. For example, the severe illness or fatality risk for a person infected by SARS-COV-2 varies widely with age and underlying health conditions\cite{levin2020assessing,hoffmann2021older}. There is a lot of empirical evidence that people are often motivated by fairness considerations\cite{fehr1999theory,fowler2005egalitarian}. That is, people are often willing to sacrifice some of their own payoff to achieve a more equitable outcome. When it comes to health inequalities, people are often very inequality averse\cite{robson2017eliciting}. Especially if an agent has vulnerable relatives, it is rather natural for her to alter her behavior during an epidemic outbreak to protect them. In the context of the current COVID-19 crisis, it has been observed that communication strategies that aim to indicate the effects of social distancing behavior on others, especially on vulnerable persons (strategy of the identifiable victim), are very effective\cite{lunn2020motivating}.

In this work, we employ a novel approach to model the agents' possible desire to keep the inequality among their payoffs below a certain threshold. Particularly, we consider that the players share a common constraint bounding inequality, modeled as their costs variance. This modeling approach can be useful for future waves of COVID-19 pandemic (probably involving variants of the virus) or for future epidemics.


Following the literature\cite{Reluga3,Poletti2,Poletti3}, we consider a compartmental model (SIR) for the spread of an epidemic and a social distancing game among the agents. The payoffs of the agents consist of two terms: a cost for the social distancing and a cost proportional to the probability of getting infected. There are two types of agents,  non-vulnerable agents, who have a small cost if they get infected and vulnerable agents, who have a higher cost. The size of the society is considered large, so the game has a continuum of players (it is a non-atomic game). In this game, the agents determine their actions to optimize their payoff and simultaneously respect a constraint concerning the variance of all the agents' payoffs.  Due to this constraint, the game is, in fact, a generalized game with a non-convex constraint. The majority of the bibliography on generalized games\cite{Kaznow,facchinei2007finite}  does not analyze generalized games with non-convex constraints. Furthermore, there are a few references on generalized non-atomic games\cite{paccagnan2018nash,jacquot2018nonsmooth,jacquot2019nonatomic}. However, in these papers, the convexity of the constraint set is built into the definition of the   Generalized Nash Equilibrium (GNE). Another related paper by Singh and Wiszniewska-Matyszkiel\cite{singh2018linear} examines a dynamic game with a continuum of players having state-dependent constraints. In this work, we give a new definition and characterization of  GNE for constrained non-atomic games. 

Numerical examples indicate that there may be many Nash equilibria inducing different costs for the players. Thus, even in the absence of social distancing regulations, it is beneficial for the players to coordinate and choose the `best' equilibrium.
In the variance constrained case, we numerically find that the inequality constraint (bounding the variance) is always beneficial for the vulnerable players. Sometimes, inequality constraints are beneficial for the group of non-vulnerable agents as well.

The rest of the paper is organized as follows. Section \ref{s.model} describes the compartmental model for the epidemic outbreak and the social distancing game between the agents. In Section \ref{s.Nash}, we analyze the game and characterize its Nash equilibria. In Section \ref{s.variance}, we introduce the constraint that concerns the variance of the payoffs and derive an appropriate definition of generalized Nash equilibrium for variance constrained games. In section \ref{s.simulations}, we present a methodology for the computation of the Nash equilibrium strategies of section \ref{s.Nash} and then give some numerical examples.   Moreover, in the same section, we present an example of generalized Nash equilibrium computation for the variance constrained game. In the Appendix, we collocate the proofs of several propositions of the paper.

\section{Mathematical Model}\label{s.model}

This section presents a variation of a popular epidemics model which assumes a continuum of agents. The state of each agent could be Susceptible (S), Infected (I), Recovered (R) or Dead (D). A susceptible person may be infected at a rate proportional to the rate she meets with infected people. Infected persons either recover or die at a constant rate. We assume that an individual recovered from the infection is immune i.e., she could not be infected again.

\begin{figure}[h!]
	\centering
	\includegraphics[width=0.20\textwidth]{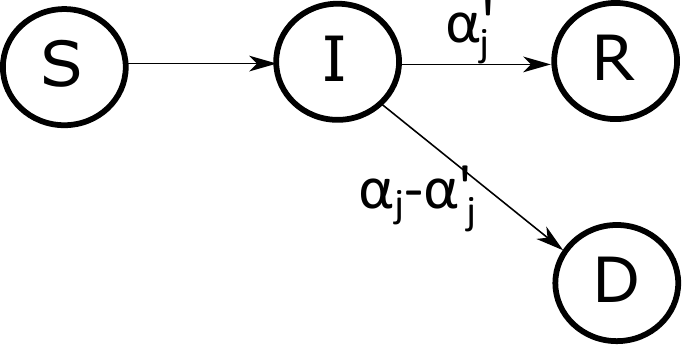}
	\caption{The Markov process describing the evolution of the state of each individual.}
	\label{markov_proc}
\end{figure}

We distinguish between two types of agents: non-vulnerable and vulnerable. We use the index $j=1$ for non-vulnerable agents and $j=2$ for vulnerable. The difference of the two types of agents is the severity of a possible infection (including the  probability to survive). An infected agent recovers with a probability rate $\alpha_j'$ and dies with a probability rate $\alpha_j-\alpha_j'$. The evolution of the individual states is presented in Figure \ref{markov_proc}.

We analyze the behavior of the agents for a time interval $[0,T]$. Denote by $u_i\in[u_m,u_M]$ the action of player $i$, indicating the fraction of time this person spends in public places. The minimum value of the actions $u_m$  describes the minimum contact a person needs for surviving and $u_M$ describes a restriction placed by the government. In the absence of a restriction we consider $u_M=1$.

\textbf{Assumption 1}: The actions  $u_i$ of all the players are \textit{ constant} during a time interval $[0,T]$. This interval represents a wave of the epidemic.
\begin{remark}
 In a more general model $u_i$ could be function of the state variables or time, but there are some arguments in favor of this choice. First, there may be a high uncertainty for the values of the state variables. Second, $u_i$'s reflect some everyday routine choices of the people and these choices may be difficult to adapt constantly.  
\end{remark}

Denote by $\mu_1$ a (Borel) measure on $[u_m,u_M]$ describing the distribution of the actions of the players  in category $1$ i.e., for $A\subset [u_m,u_M]$, the value of $\mu_1(A)$ denotes the mass of the players using an action $u\in A$.  Similarly, denote by  $\mu_2 $ the distribution of actions of the players of type $2$. The total mass of players of types $1$ and $2$ is  $n_1$ and $n_2$ respectively i.e., $\mu_1([u_m,u_M])=n_1$ and $\mu_2([u_m,u_M])=n_2$. We first describe the evolution of the epidemic for general distributions $\mu_1,\mu_2$. \par
Denote by $  S_{1u}(t)$ the probability a non-vulnerable player who plays $u\in [u_m,u_M]$ to be susceptible at time $t$ and by $  I_{1u}(t)$ the probability to be infected. Similarly define  $  S_{2u}(t)$,  $ I_{2u}(t)$. The rate at which this  person gets infected is given by: $ruI^f,$ where $r$ is a positive constant and $I^f$  denotes the density of infected people in `public places'. Each player contributes to $I^f(t)$ proportionally to her probability of being infected at time $t$. The dynamics is given by:
\begin{equation}
\begin{aligned}
\dot { S}_{ju} &= -ru  S_{ju}I^f(t)\\
\dot { I}_{ju} &=ru S_{ju}I^f(t)-\alpha_1  I_{ju}\\
\dot {z}& = I^f(t)
\label{meas_DE}
\end{aligned},
\end{equation}
where  $z$ is an auxiliary variable, $j=1,2$ and:
\begin{equation}\label{free_inf}
I^f(t)=\int_{[u_m,u_M]}  I_{1u'}(t)u' \cdot\mu_1(du')+ \int_{[u_m,u_M]}  I_{2u'}(t)u' \cdot\mu_2(du').
\end{equation}
The initial conditions are
\begin{equation}
S_{1u}(0)= S_{2u}(0)= (1-I_0),  ~    I_{1u}(0)= I_{2u}(0)=I_0
\label{Gen_init_cond},
\end{equation}
where $I_0$ is the percentage of infected persons at time $0$. Here  we assume, without loss of generality, that at the beginning of the time interval $[0,T]$, the agents of both types are infected with the same probability $I_0$.  

Before showing the existence of a solution for the initial value problem  \eqref{meas_DE}--\eqref{Gen_init_cond}, we introduce some function spaces. Let:
$$X=C([u_m,u_M],\mathbb R^4)\times \mathbb R,$$
where $C([u_m,u_M],\mathbb R^4)$ be the space of continuous functions defined on $[u_m,u_M]$ and values on $\mathbb R^4$. The space $X$ equipped with the norm:
$$\|x\| = \max\{|x_{1u}|,|x_{2u}|,|x_{3u}|,|x_{4u}|,|x_5|: u\in[u_m,u_M] \},$$
is a Banach space. We also consider the Banach space $Y$ of signed measures on $[u_m,u_M]$ with the total variation norm.

\begin{proposition}\label{thm1}
	The initial value problem \eqref{meas_DE}--\eqref{Gen_init_cond} has a unique solution. Furthermore, this solution is continuous on $u$.
\end{proposition}
\textit{Proof:} See  Appendix \ref{PropExPr}. \hfill $\square$

The cost  of an agent $i$ of type $j$ consists of two terms. The first term is proportional to the probability of getting infected and the vulnerability of the player. The second term, represents the benefits earned from social interactions. The cost is given by:
\begin{equation}
J_i = G_j P_i -Q_j(u_i,\bar u_1,\bar u_2),
\label{Ind_cost}
\end{equation}
where $G_j $ 
corresponds to the expected severity of a possible infection, $P_i$ is the probability that 
$i$ gets infected within the time interval $[0,T]$.  Note that $G_2>G_1$. The quantity $Q_j(u_i,\bar u_1,\bar u_2)$ represents the  utility derived from the interaction with others where $ \bar u_1,\bar u_2$  are the mean actions of the players of types $j=1$ and $j=2$ respectively.
For simplicity we assume that $Q_j$ has the form:
	$$Q_j(u_i,\bar u_1,\bar u_2)=s_{j1}u_i\bar u_1+s_{j2}u_i\bar u_2,$$
where $s_{j1},s_{j2}$ are non-negative constants.
\begin{remark}
	In the computation of the  second term in \eqref{Ind_cost}, we assume that the number of people in these types is approximately constant with time. This  is a good approximation in epidemics with a low mortality rate and duration small compared to the average human life. 
\end{remark}

Let us then compute the probability of getting infected $P_i$. The probability ${S}^i_{ju_i}(t)$ that an agent $i$ of either type ($j=1$ or $j=2$) is not infected up to time $t$ evolves according to:
$$\dot{{S}}^i_{ju_i}=- r {S}^i_{ju_i} u_i I^f.$$
Thus, we have:
$$P_i= 1-{S}^i_{ju_i}(T) = I_0+ (1-I_0) \left[1-\exp\left(  -r u_i\int_0^T I^f(t)dt \right)\right].$$
Here we assume that at the beginning of the time interval $[0,T]$, all the agents are infected with a small probability $I_0$. Denoting by $F(\mu_1,\mu_2)=r\int_0^T I^f(t)dt$ the cost is written as:
$$J_j(u_i,\mu_1,\mu_2) = G_j \left[I_0+ (1-I_0) \left[1-\exp\left(  -u_i F(\mu_1,\mu_2) \right)\right]\right]-s_{j1}u_i\bar u_1-s_{j2}u_i\bar u_2.$$
Since we assume a very large population of players, each one of them is not able to affect the distributions $\mu_1, \mu_2$. It is interesting to observe that the individual cost $J_i$ is concave in $u_i$. To see this take the second derivative of $J_i$ with respect to $u_i$:
$$\frac{\partial^2 J_i}{\partial u_i^2} =- G_j(1-I_0) \exp\left(  -r u_i\int_0^T I^f(t)dt \right) \left(  -\int_0^T I^f(t)dt \right) ^2<0.$$
Therefore, the possible actions minimizing the individual cost are $u_i=u_m$ and $u_i=u_M$.
Thus, to compute the Nash equilibria, we focus on distributions assigning the entire mass  on $\{u_m,u_M\}$.

\begin{remark}
	\label{ConcavRemark}
	The fact that the cost function $J_i$ is concave in $u_i$ simplifies the analysis a lot, implying that  players choose either $u=u_m$ or $u=u_M$. It further allows us to describe dynamics using a finite-dimensional model. 
\end{remark}


\section{Nash Equilibrium}\label{s.Nash}
To analyze the Nash equilibrium we focus on distributions having all the mass on $\{u_m,u_M\}$.The dynamics is given by:
\begin{equation}
\begin{aligned}
\dot S_{ju_m} &= -ru_mI^fS_{ju_m}, &
\dot S_{1u_M} &= -ru_MI^fS_{ju_M}, \\
\dot I_{ju_m} &=  ru_mI^fS_{ju_m}- \alpha_j I_{ju_m},~~ &
\dot I_{ju_M} &=  ru_MI^fS_{ju_M}- \alpha_j I_{ju_M}, \\
\dot z&=I^f,
\end{aligned} \label{SIR_diff_eq}
\end{equation}
where $j=1,2$, the total mass of `free infected people' $I^f$ is given by:
$$I^f(t) = \sum_{j=1}^2n_j((1-\tilde u_j)u_mI_{ju_m}(t)+ \tilde u_ju_MI_{ju_M}(t)),$$ and
$\tilde u_j=\mu_j(\{u_M\})$ is the percentage of players of type $j$ using $u_M$. The initial conditions are given by:
\begin{equation}
\begin{aligned}
S_{ju_m}(0) = S_{ju_M}(0)  =   (1-I_0) , ~~~
I_{ju_m} (0)=    I_{ju_m} (0)= I_0 ,~~~
z(0)=0
\end{aligned}
\label{Init_cond}
\end{equation}

Denote by $\phi_{\tilde u_1,\tilde u_2}^z(t)$ the $z$-part of the solution of the  differential equation \eqref{SIR_diff_eq} with initial conditions \eqref{Init_cond}. Then it holds:
$$F(\mu_1,\mu_2)=F(\tilde u_1,\tilde u_2) =  r\phi^z_{\tilde u_1,\tilde u_2}(T).$$

\begin{remark}
	We view the equilibria where some players of type $j$ play $u_m$ and some $u_M$ as equilibria in \emph{symmetric  mixed} strategies. Particularly, each player of type $j$ plays $u_M$ with probability $\tilde u_j$.
\end{remark}

\begin{proposition}
	\label{NEchar}
	Consider a set of strategies characterized by $\tilde u_1,\tilde u_2$, where a fraction $1-\tilde u_j$ of the players of type $j$ use $u=u_m$ and a fraction $\tilde u_j$ of the players of type $j$ use $u=u_M$. This set of strategies is a Nash equilibrium if and only if, for each $j=1,2$, one of the following holds:
	\begin{itemize}
		\item[(i)] $0<\tilde u_j<1$ and:
		$$G_j(1-I_0) (e^{-u_mF(\tilde u_1,\tilde u_2) }-e^{-u_MF(\tilde u_1,\tilde u_2) })  =   (u_M-u_m)(s_{j1}\bar u_1+s_{j2}\bar u_2),$$
		where $\bar u_j = u_m+(u_M-u_m)\tilde u_j.$
		\item[(ii)] $\tilde u_j=0$ and:
		$$G_j(1-I_0) (e^{-u_mF(\tilde u_1,\tilde u_2) }-e^{-u_MF(\tilde u_1,\tilde u_2) })  \geq   (u_M-u_m)(s_{j1}\bar u_1+s_{j2}\bar u_2).$$
		\item[(iii)] $\tilde u_j=1$ and:
		$$G_j(1-I_0) (e^{-u_mF(\tilde u_1,\tilde u_2) }-e^{-u_MF(\tilde u_1,\tilde u_2) })  \leq   (u_M-u_m)(s_{j1}\bar u_1+s_{j2}\bar u_2).$$
	\end{itemize}
\end{proposition}
\textit{Proof}:  The proof is immediate, 
observing that (i) corresponds to the case where the players of type $j$ are indifferent between $u_m$ and $u_M$ and (ii), (iii)  correspond to preference of $u_m$ over $u_M$ and $u_M$ over $u_m$ respectively. \hfill $\square$

The existence of a Nash equilibrium is a consequence of Theorem 1 of  Mas-Colell(1984)\cite{mas1984theorem}.

\begin{corollary}
	\label{COrStr}
	Assume that   $s_{11}=s_{21}$ and $s_{12}=s_{22}$. Then the possible Nash equilibria $ (\tilde u_1,\tilde u_2)$ are in of one of the following forms 
	$(0,0),~ (\tilde u_1,0),~ (1,0),~ (1,\tilde u_2),~ (1,1)$. 
\end{corollary}
\textit{Proof}:  Let $ (\tilde u_1,\tilde u_2)$ be a Nash equilibrium. Then, if $\tilde u_1<1$  it holds:
$$G_1(1-I_0) (e^{-u_mF(\tilde u_1,\tilde u_2) }-e^{-u_MF(\tilde u_1,\tilde u_2) })  \geq   (u_M-u_m)(s_{11}\bar u_1+s_{12}\bar u_2).$$
Since,  $s_{11}=s_{21}$, $s_{12}=s_{22}$ and $G_2>G_1$ we have:
$$G_2(1-I_0) (e^{-u_mF(\tilde u_1,\tilde u_2) }-e^{-u_MF(\tilde u_1,\tilde u_2) })  >   (u_M-u_m)(s_{21}\bar u_1+s_{22}\bar u_2).$$
But, since $ (\tilde u_1,\tilde u_2)$ is a Nash equilibrium Proposition \eqref{NEchar} implies that $\tilde u_2=0$ \hfill $\square$
\begin{corollary}
	If $u_m=0$, then $ (\tilde u_1,\tilde u_2)=(0,0)$ is always a Nash equilibrium. 
\end{corollary}

\section{The Variance Constrained Game}
\label{s.variance}

This section analyzes a game situation, where the players pose a shared bound on the variance of their costs. To do so, we first introduce a notion of equilibrium with a shared constraint, for non-atomic games, and then characterize it in terms of small variations. 
We assume  that the strategies of the players are symmetric (that is all the players of the same type use the same strategy), allowing for randomization. Consider a pair of  distributions $(\mu_1,\mu_2)$ for the actions of the players. Then, the players of type $j$ randomize according to $\bar \mu_j(\cdot)=\mu_j(\cdot)/n_j$.

The variance of the costs is given by:
\begin{equation}
\begin{aligned}
V(\mu_1,\mu_2) &= \frac{ \int  (J_1(u',\mu_1,\mu_2)-\bar J)^2\mu_1(du')+\int  (J_2(u',\mu_1,\mu_2)-\bar J)^2\mu_2(du')} {n_1+n_2},\\
&= \frac{ n_1\int  (J_1(u',\mu_1,\mu_2)-\bar J)^2\bar \mu_1(du')+n_2\int  (J_2(u',\mu_1,\mu_2)-\bar J)^2\bar \mu_2(du')} {n_1+n_2},\\
\end{aligned}
\end{equation}
where
$\bar J=(n_1 \bar J_1+n_2 \bar J_2 )/(n_1+n_2)$, and:
$$\bar J_j=\frac{1}{n_j}\int  J_j(u',\mu_1,\mu_2) \mu_j(du')=\int  J_j(u',\mu_1,\mu_2) \bar\mu_j(du').$$

We then  describe a notion of equilibrium for the generalized game with variance constraint. Ideally, to have an equilibrium, the actions of each player should minimize the cost subject to the variance constraint.  The difficult point here is that, since we have a continuum of players, the variance does not depend on the actions of individual players.
To define a meaningful notion of equilibrium, instead of analyzing the effect of a deviation of a single player, we consider the deviation of a small fraction of players of type $j$  and see how a variation from the nominal mixed strategy $\bar \mu_j$
affects the cost of this small group of players and the total variance.
Then, we take the limit as the total mass of the group of players tends to zero.

Denote by $j$ the type of players containing the deviating group and by $-j$ the other type of players. Assume that the total mass of deviating players is $\varepsilon$ and that the deviating players use a mixed strategy $\bar\mu_j'$ (note that it holds $\bar\mu_j'([u_m,u_M])=1$). Then, the distribution of the actions of the players of type $j$ is given by: $$\mu_j+ \varepsilon(\bar \mu'_j-\bar \mu_j)=\mu_j+ \varepsilon \delta  \mu_j. $$

The mean cost of the deviating players after the deviation is:
$$\bar J_j^{\text{dev}}= \int  J_j(u',\mu_j+\varepsilon(\bar \mu'_j-\bar \mu_j),\mu_{-j}) {\mu_j}'(du'),$$
while before the deviation is $\bar J_j$. The following lemma expresses this the limit of this deviation, as well as the directional (Gateaux) derivative of the variance, in terms of linear bounded operators. This result will be used to define the Generalized Nash equilibrium.

\begin{lemma}For all $\mu_{-j}$ it holds:
	\begin{itemize}\item[(i)]
		The limit of the variation $\bar J_j^{\text{dev}}-\bar J_j$, as $\varepsilon\rightarrow0$, is a linear function of $\delta \mu_j$. Particularly, it is written as:
		\begin{equation}
		\lim_{\varepsilon\rightarrow0}(\bar J_j^{\text{dev}}-\bar J_j)= \mathcal{K}^j_{\mu_1,\mu_{2}} \delta \mu_j=\int  J_j(u',\mu_j,\mu_{-j}) {\delta\mu_j}(du'),
		\label{Cost_variation}
		\end{equation}
		where   $\mathcal{K}^j_{\mu_1,\mu_{2}} \in Y^\star$ and $Y^\star$ is the space of bounded linear functionals on $Y$ .
		\item[(ii)]
		The directional derivative of the variance $V(\mu_j,\mu_{-j})$ in the direction $\delta\mu_j$  is expressed as:
		$$\lim_{\varepsilon\rightarrow0}\frac{V(\mu_j+\varepsilon\delta\mu_j,\mu_{-j})-V(\mu_j,\mu_{-j}) }{\varepsilon}=\mathcal{L}^j_{\mu_1,\mu_2}\delta\mu_j,$$
		where $\mathcal{L}^j_{\mu_1,\mu_2}\in Y^\star$. Furthermore,   $\mathcal{L}^j_{\mu_1,\mu_2} $  can be written as:
		$$
		\mathcal{L}^j_{\mu_1,\mu_2}\delta\mu_j=
		\int f^\text{var}_{j,V,\mu_1,\mu_2}(u')\delta\mu_j(du'), $$
		with $f^\text{var}_{j,V,\mu_1,\mu_2}(u')$ continuous.
	\end{itemize}
	\label{Sect_3_lemma}
\end{lemma}
\textit{Proof:} See  Appendix \ref{Sect_3_lemma_Pr}. \hfill $\square$

\begin{definition}[Generalized Nash Equilibrium]
	\label{GNE_DEF}
	A distribution of actions described by $(\mu_1,\mu_2)$ is a Generalized Nash Equilibrium (GNE) with variance constraint $V \leq C$ if either:
	\begin{itemize}
		\item[(i)] $V(\mu_1,\mu_2)<C$ and for any $j=1,2$ and any probability measure $\bar\mu'_j$, it holds:
		$$\mathcal{K}^j_{\mu_1,\mu_{2}} \delta \mu_j\geq 0,$$
		where $\delta \mu_j = \bar\mu'_j-\bar \mu_j$, or
		\item[(ii)] $V(\mu_1,\mu_2)=C$ and for any $j=1,2$ and any probability measure $\bar \mu'_j$, it holds:
		$$\mathcal{K}^j_{\mu_1,\mu_{2}} \delta \mu_j< 0\implies \mathcal{L}^j_{\mu_1,\mu_2} \delta \mu_j> 0,$$
		where $\delta \mu_j = \bar\mu'_j-\bar \mu_j$.
	\end{itemize}
\end{definition}

\begin{remark}
	In the first case of the definition, any small group of players is not sufficient to increase the variance above $C$. Thus, if $\mu_1,\mu_2$ is an equilibrium, there is no profitable deviation, and the definition coincides with the equilibrium of Section \ref{s.Nash}.  In the second case, $\mu_1,\mu_2$ is an equilibrium, if any  profitable deviation for a small group of players, increases the variance above $C$.
\end{remark}

\begin{remark}
	Some  notions of GNE for games with a continuum of players were already introduced   in the literature\cite{paccagnan2018nash,jacquot2018nonsmooth,jacquot2019nonatomic}. However, these definitions assume a convex constraint set. Let us note that Definition \ref{GNE_DEF} is neither a generalization nor a special case of the these definitions.
\end{remark}

We then introduce a refinement of GNE, called non-singular GNE. It turns out that  non-singular GNE are easier to compute.
\begin{definition}[non-singular GNE]
	A pair $(\mu_1,\mu_2)$ is variance stationary if either:
	\begin{itemize}
		\item[(i)] for all $\delta \mu_1 = \bar\mu'_1-\bar \mu_1$, with $ \bar\mu'_1$ probability measure, it holds:
		$$\mathcal{L}^1_{\mu_1,\mu_2} \delta \mu_1\geq 0,$$
		or
		\item[(ii)] for all $\delta \mu_2 = \bar\mu'_2-\bar \mu_2$, with $\bar \mu'_2$ probability measure, it holds: $$\mathcal{L}^2_{\mu_1,\mu_2} \delta \mu_2\geq 0.$$
	\end{itemize}
	We call a GNE $(\mu_1,\mu_2)$  non-singular if it is not  variance stationary.
\end{definition}

\begin{lemma}
	Assume that $(\mu_1,\mu_2)$ is not variance stationary. Then, $(\mu_1,\mu_2)$ is a Generalized Nash equilibrium with variance constraint $V \leq C$ if and only if either it satisfies (i) of Definition \ref{GNE_DEF} or
	$V(\mu_1,\mu_2)=C$ and for any probability measure $\bar\mu'$, it holds:
	\begin{equation} \label{GNE_CHar}
	\mathcal{K}^j_{\mu_1,\mu_{2}} \delta \mu_j< 0\implies \mathcal{L}^j_{\mu_1,\mu_2} \delta \mu_j\geq 0,\end{equation}
	where $\delta \mu_j = \bar\mu'_j-\bar \mu_j$.
\end{lemma}
\textit{Proof:} The direct part is immediate. Assume that the converse is not true, that is, there is a probability measure $\bar\mu'_j$ such that:
$$\mathcal{K}^j_{\mu_1,\mu_{2}}  (\bar\mu'_j-\bar \mu_j)< 0,~ \mathcal{L}^j_{\mu_1,\mu_2}  (\bar\mu'_j-\bar \mu_j)= 0.$$
Then, since $(\mu_1,\mu_2)$ is not variance stationary, there is a probability measure $\bar\mu''_j$ such that
$\mathcal{L}^j_{\mu_1,\mu_2} (\bar\mu''_j-\bar \mu_j)<0$. Hence, there is a $\theta\in(0,1)$ such that:
$$\mathcal{K}^j_{\mu_1,\mu_{2}}  (\theta\bar\mu'_j+(1-\theta)\bar\mu''_j-\bar \mu_j)< 0,~ \mathcal{L}^j_{\mu_1,\mu_2}  (\theta\bar\mu'_j+(1-\theta)\bar\mu''_j-\bar \mu_j)< 0.$$
But this contradicts \eqref{GNE_CHar}. \hfill $\square$

Assume that $V_A(\mu_1,\mu_2)=C$ and $(\mu_1,\mu_2)$ is not variance stationary. Then, $(\mu_1,\mu_2)$ is an equilibrium if and only if   there is no $\delta \mu_j=\bar\mu'_j-\bar\mu_j$ such that:
\begin{equation}
\mathcal{K}^j_{\mu_1,\mu_{2}}  \delta \mu_j< 0 \text{~ and ~} \mathcal{L}^j_{\mu_1,\mu_{2}} \delta \mu_j< 0. \label{Ineq1}
\end{equation}
The following proposition characterizes the non-singular GNE in terms of measures supported on at most two points.

\begin{proposition}
	If there is a probability measure $\bar\mu'_j$ such that \eqref{Ineq1} holds true, then there is another probability measure $\bar\mu''_j$ supported on at most two points which also satisfies \eqref{Ineq1} with  $\delta \mu=\bar\mu''-\bar\mu$.
	\label{TwoPointProp}
\end{proposition}
\textit{Proof:} See  Appendix \ref{TwoPointPropPr} \hfill $\square$

Let us introduce the following quantities:  $$g^{\mathcal K}_{j,\mu_1,\mu_2}(u) = \mathcal{K}^j_{\mu_1,\mu_{2}}  (\mathcal d_u-\bar\mu_j), ~~~~g^{\mathcal L}_{j,\mu_1,\mu_2}(u) = \mathcal{L}^j_{\mu_1,\mu_{2}}  (\mathcal d_u-\bar\mu_j),$$ where $\mathcal d_u$ is a Dirac measure supported on $u$.
Using these quantities we have the following necessary (Corollary \ref{CorollNec}), and necessary  and sufficient conditions (Corollary \ref{Coroll2}).
\begin{corollary}
	\label{CorollNec}
	If $(\mu_1,\mu_2)$ is a GNE then for all $u\in[u_m,u_M]$, $j=1,2$ if $g^{\mathcal K}_{j,\mu_1,\mu_2}(u)<0$ then  $g^{\mathcal L}_{j,\mu_1,\mu_2}(u)\geq 0$.
\end{corollary}
\begin{corollary}
	\label{Coroll2}
	A non variance stationary pair $(\mu_1,\mu_2)$ is a GNE if and only if for all $u',u''\in[u_m,u_M]$, $\rho\in[0,1]$, $j=1,2$ if $\rho g^{\mathcal K}_{j,\mu_1,\mu_2}(u')+(1-\rho) g^{\mathcal K}_{j,\mu_1,\mu_2}(u'')<0$ then  $\rho g^{\mathcal L}_{j,\mu_1,\mu_2}(u')+(1-\rho) g^{\mathcal L}_{j,\mu_1,\mu_2}(u'')\geq 0$.
\end{corollary}

\begin{remark}
The proposed formulation describes pro-social behaviors in terms of bounding the variance of the costs. There are various alternative formulations. For example, people may bound the maximum number of infected individuals, reflecting the bounded capacity of the healthcare systems. Another alternative would be to consider altruistic players\cite{shim2012influence}. Finally, pro-social behavior can be modeled as Kantian behavior\cite{kordonis2020model}.  We chose to model pro-social behavior as bounding the variance, because of the vast health inequities created by the current COVID-19 pandemic.
\end{remark}

\section{Computational Study}
\label{s.simulations}
We then present some numerical results. In Subsection \ref{NE_COmp_Sect}, we compute numerically the Nash equilibria  of the unconstrained game providing two illustrative examples and in Subsection \ref{ConstrNumSubSect},  we study an example for the variance constrained game.
\subsection{Computing Unconstrained Nash Equilibria}
\label{NE_COmp_Sect}
The computation of the value of $F(\bar u_1,\bar u_2)$ corresponds to the numerical integration of \eqref{SIR_diff_eq}.
The search for pure Nash equilibria needs just  the computation of $F(0,1), F(1,0)$, and $F(1,1)$ and checking the corresponding inequalities.

Let us then describe the procedure to find equilibria in the form of the  $(\tilde u_1,0)$ or $(\tilde u_1,1)$. We have first to find the solutions of:
\begin{align*}
H_{\tilde u_2}(\tilde u_1)=G_1(1-I_0)  (e^{-u_mF(\tilde u_1,\tilde u_2) }-e^{-u_MF(\tilde u_1,\tilde u_2) })-     (u_M-u_m)(s_{j1}(u_m+(u_M-u_m)\tilde u_1)+s_{j2}\bar u_2)=0,
\end{align*}
with respect to $\tilde u_1$, for a fixed value of $\tilde u_2=0,1$.
To do so we  use line search (an alternative, would be to use a multi-start Newton algorithm).  
Having found a solution of $H_{\tilde u_2}(\tilde u_1)=0$ for $\tilde u_2=0$ or $\tilde u_2=1$ we need also to check the corresponding inequality. 
The computation of possible equilibria in the form $(0,\tilde u_2)$ or $(1,\tilde u_2)$ is similar.

Let us compute any possible Nash equilibrium where both types use mixed strategies (internal Nash equilibria).  Then we should have:
\begin{equation}
\begin{aligned}
G_1(1-I_0) (e^{-u_mF(\tilde u_1,\tilde u_2) }-e^{-u_MF(\tilde u_1,\tilde u_2) })  &=   (u_M-u_m)(s_{11}\bar u_1+s_{12}\bar u_2)\\
G_2(1-I_0) (e^{-u_mF(\tilde u_1,\tilde u_2) }-e^{-u_MF(\tilde u_1,\tilde u_2) })  &=   (u_M-u_m)(s_{21}\bar u_1+s_{22}\bar u_2)
\end{aligned}\label{Int_NE}
\end{equation}
Any solution of this equation should belong to the line:
$$G_2   (s_{11}\bar u_1+s_{12}\bar u_2)=G_1  (s_{21}\bar u_1+s_{22}\bar u_2)$$
or equivalently:
\begin{equation}
(G_2 s_{11} -G_1s_{21})\tilde u_1+(G_2s_{12}-G_1s_{22})\tilde u_2 = \frac{G_1(s_{21}+s_{22})-G_2(s_{11}+s_{12})}{u_M-u_m}u_m.
\label{inter_line}
\end{equation}

Therefore,  to find any internal Nash equilibria, we  examine using line search if there are solutions of \eqref{Int_NE} on the line  \eqref{inter_line}.
We then present some examples with concrete values for the parameters.

\textbf{Example 1: }
The parameters are $T=100$, $r=5/16$, $I_0=0.01$, $ n_1=0.8$, $n_2=0.2$,  $\alpha _1=\alpha_2=1/8$. These parameters correspond to an epidemic with basic reproduction number $R_0=2.5$, where people remain infectious for a mean time of 8 days \cite{kabir2020evolutionary}. The $s$ parameters are $s_{11}=s_{21}=2$ and $s_{12}=s_{22}=0.5$. We assume that $u_M=0.8$ and $u_m=0.5$. We compute the equilibria for different values of $G_1$ and $G_2$, assuming that $G_2/G_1=10$.
This choice roughly corresponds to the infection fatality risks of the older people compared with  the infection fatality risks of younger people\cite{levin2020assessing}. The variation described corresponds to varied ways that people may weight health, money and well being. 

\begin{figure}[h!]
	\centering
	\includegraphics[width= 0.8\textwidth]{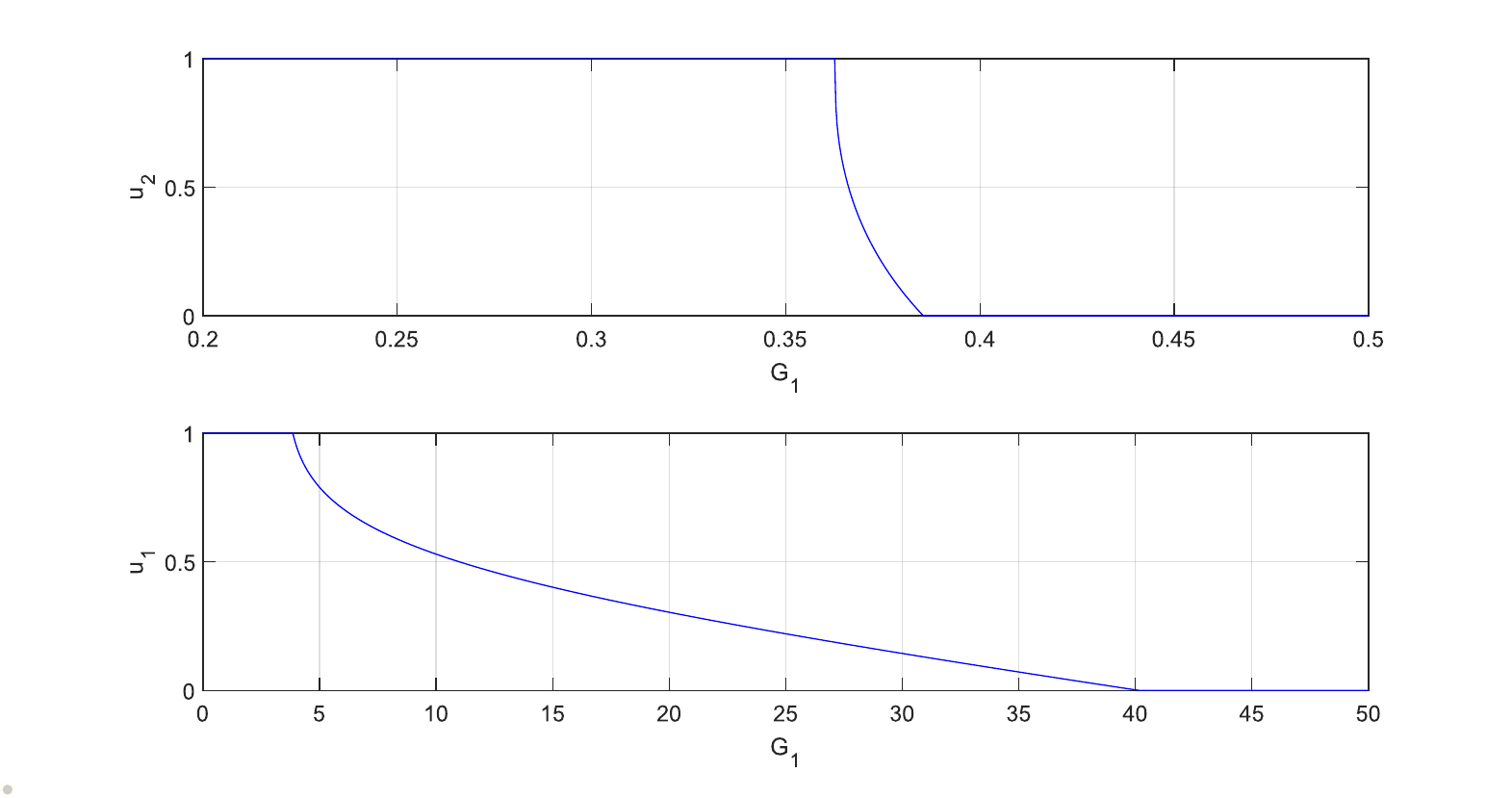}
	\caption{The upper part of the image presents the equilibria when $G_1\in[0.2,0.5]$. In this region, players of type 1 (non-vulnerable players) play $u=u_M$ and players of type 2 randomize between $u_m$ and $u_M$. The probability to play $u_M$ denoted by $\tilde u_2$ is illustrated in the upper part. Similarly the lower part presents the equilibria for $G_1\in[0.5,50]$. In this region, all the players of type 2 play $u=u_m$, while the players of type 1 randomize with probability $\tilde u_1$.  }
	\label{Bifurcation_figure}
\end{figure}

The equilibria of the game are presented in Figure \ref{Bifurcation_figure}.  We observe for all the values of $G_1, G_2$ with $G_2/G_1=10$ there is a unique Nash equilibrium.  When $G_1,G_2$ are small, the equilibrium strategies are $\tilde u_1=\tilde u_2=1$. Then, as $G_1,G_2$ become larger, there is a mixed Nash equilibrium $(1,\tilde u_2)$.  For intermediate values of $G_1,G_2$, there is a unique equilibrium with $\tilde u_1=1$ and $\tilde u_2=0$. For larger values of $G_1,G_2$  the equilibrium has the form $\tilde u_1=\tilde u_2=0$.   Finally for large $G_1,G_2$ there is a unique equilibrium $\tilde u_1=0$ and $\tilde u_2=0$.
 
\begin{figure}[h!]
	\centering
	\includegraphics[width= 0.8\textwidth]{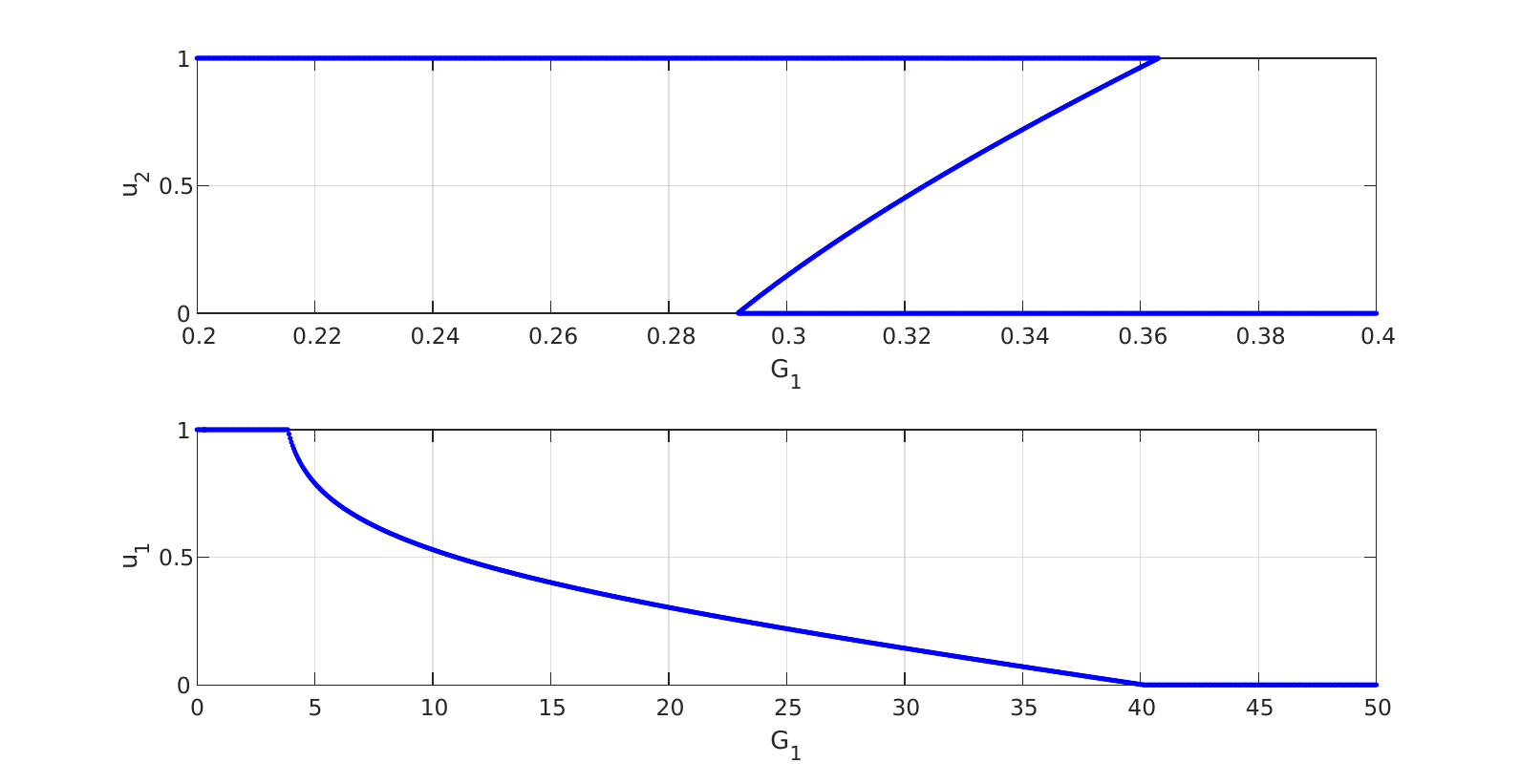}
	\caption{The figure presents the equilibria of the game. The upper part present $\tilde u_2$, when $G_1\in[0.2,0.4]$. In this interval all the players of type 1 play $u_M$, that is $\tilde u_1=1$. There are at most three equilibria. The equilibria for $G_1\in[0.4,50]$ are presented in the lower part of the figure. In this region all the players of type 2 play $u=u_m$, while the players of type 1 randomize with probability $\tilde u_1$.  }
	\label{Bifurcation_figure2}
\end{figure}

\textbf{Example 2: }
In this example there is a strong homophily. Particularly, $s_{11}=s_{22}=2$ and $s_{21}=s_{12}=0.5$. The rest of the parameters are as in Example 1, including the fact that $G_2/G_1=10$. The equilibria for various values of are presented in Figure   \ref{Bifurcation_figure2}. For low values of $G_1$ there is a unique pure Nash equilibrium, where all the players play $u_M$. Then around $G_1=0.292$, in addition to the pure equilibrium $(1,1)$, a pair of equilibria appears. One of the new equilibra is  pure  and the other  is mixed. In the new pure equilibrium all the players of type 1 play $u_M$ and all the players of type 2 play $u_m$. In the mixed equilibrium, players of type 2 randomize, that is the mixed equilibrium has the form $(1,\tilde u_2)$. As $G_1$ becomes larger the value of $\tilde u_2$ increases and eventually, around $G_1=0.363$, the mixed equilibrium $(1,\tilde u_2)$ meets  with the pure equilibrium $(1,1)$ and they both disappear. Then, for $G_1\in[0,363,3.8]$ there is a unique Nash equilibrium where all the players of type 1 play $u_M$ and players of type 2 play $u_m.$ On the interval $G_1\in[3.8, 40.1]$, there is a unique mixed Nash equilibrium where all the players of type 2 play $u_m$ and the players of type 1 randomize. For larger values of $G_1$ all the players play $u_m$.
 
\subsection{Computing Constrained Equilibria}
\label{ConstrNumSubSect}
We then search for generalized Nash equilibria with variance constraints. Let us first note that if the pair $(\mu_1,\mu_2)$ satisfies Definition \ref{GNE_DEF}.(i) then it also is an unconstrained Nash equilibrium. Thus, it is sufficient to check if the Nash equilibria computed in the previous section satisfy the constraint $V(\mu_1,\mu_2)\leq C$.

We then compute GNE, satisfying Definition \ref{GNE_DEF}.(ii), in  the case  where the policies $(\mu_1,\mu_2)$ are of the form $\mu_1 = n_1 \mathcal d_{u_1}
$, $\mu_2 = n_2 \mathcal d_{u_2}$, where $\mathcal d_u$ is a Dirac measure concentrated on $u$. We use a grid to find the pairs $(u_1,u_2)$ such that the equality $V(\mu_1,\mu_2)=C$ holds approximately. These points are candidates for GNE. For each of these points in the grid, we compute the functions $g^{\mathcal K}_{j,\mu_1,\mu_2}(u)$ and $g^{\mathcal L}_{j,\mu_1,\mu_2}(u)$ for $j=1,2$ and a grid of points $u$. The details of the computation of 
$g^{\mathcal K}_{j,\mu_1,\mu_2}(u)$ and $g^{\mathcal L}_{j,\mu_1,\mu_2}(u)$ are given in  Appendix \ref{Appendix.GNEcomp}. We then use Corollary \ref{Coroll2}  to check weather each of these points is a GNE.  

 \textbf{Example 3: }
	In this example the parameters are as in Example 1, and the vulnerability parameters $G_1=8$ and $G_2=80$.  The unconstrained Nash equilibrium is $\tilde u_1 =     0.602, \tilde u_2=0$. The cost for the non-vulnerable and vulnerable players under the Nash equilibrium are $J_1 =0.185$ and $J_2 =9.12$ respectively. 

\begin{figure}[h!]
		\centering
		\includegraphics[width=0.6\textwidth]{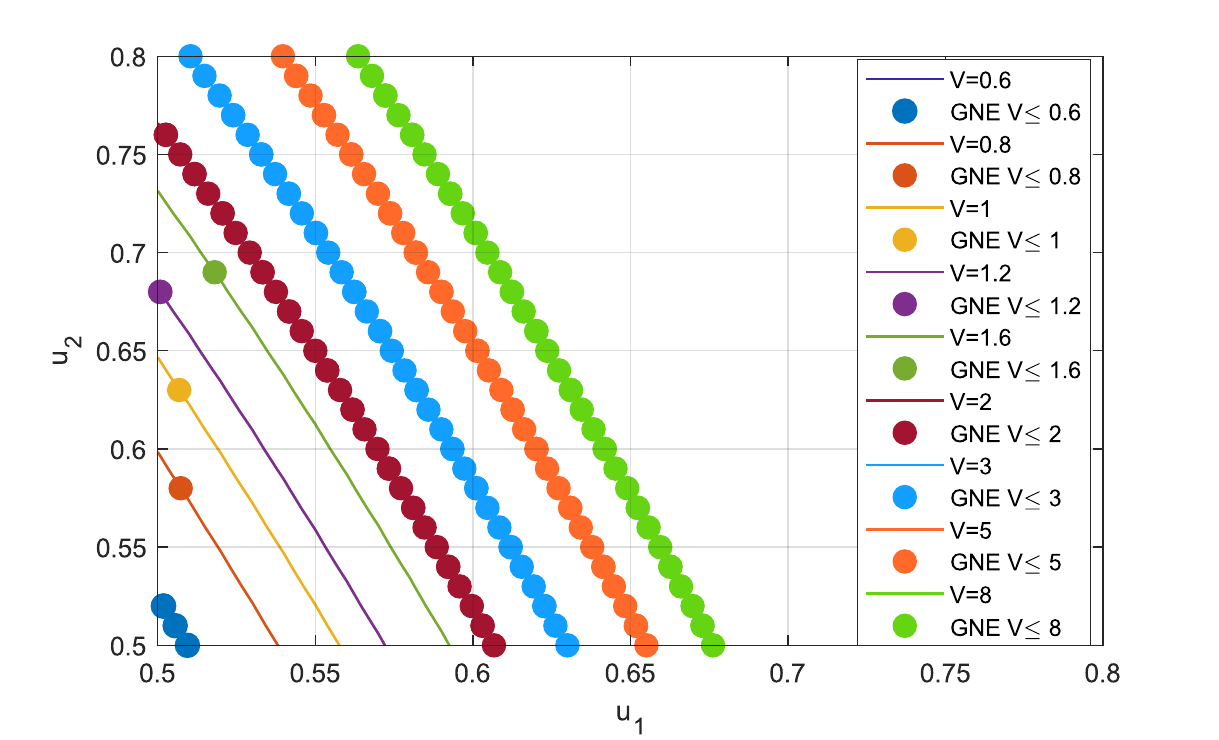}
		\caption{The contour line $V=C$, for various values of $C$ and the corresponding GNE.  }
		\label{many_GNEFig}
	\end{figure}

\begin{figure}[h!]
		\centering
		\includegraphics[width=0.8\textwidth]{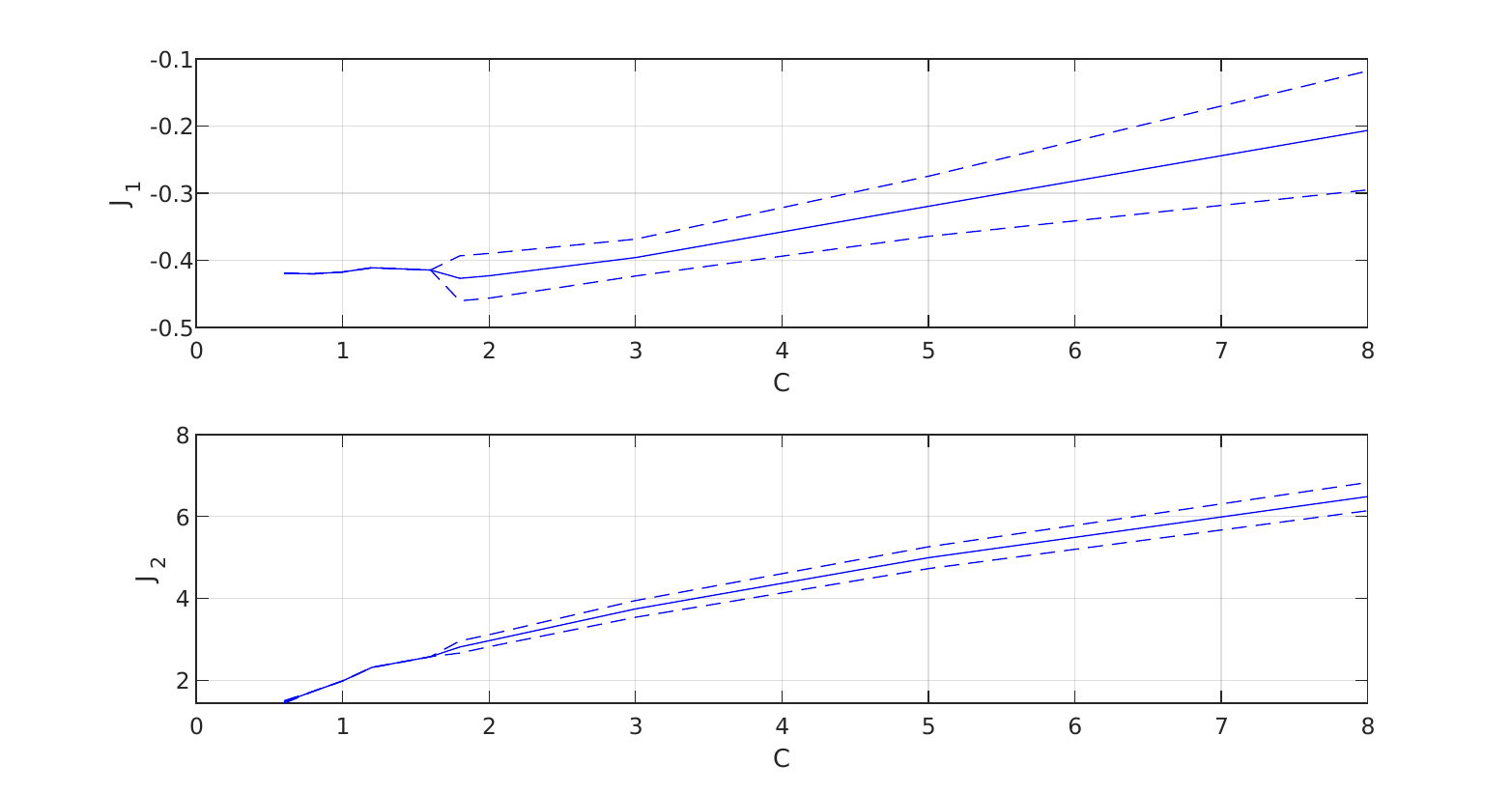}
		\caption{ The costs for vulnerable and non-vulnerable players for various values of $C$. The dashed lines correspond to the minimum and maximum values for all the equilibria and the solid lines for an average value. }
		\label{many_GNEFig1}
	\end{figure}

\begin{figure}[h!]
		\centering
		\includegraphics[width=0.8\textwidth]{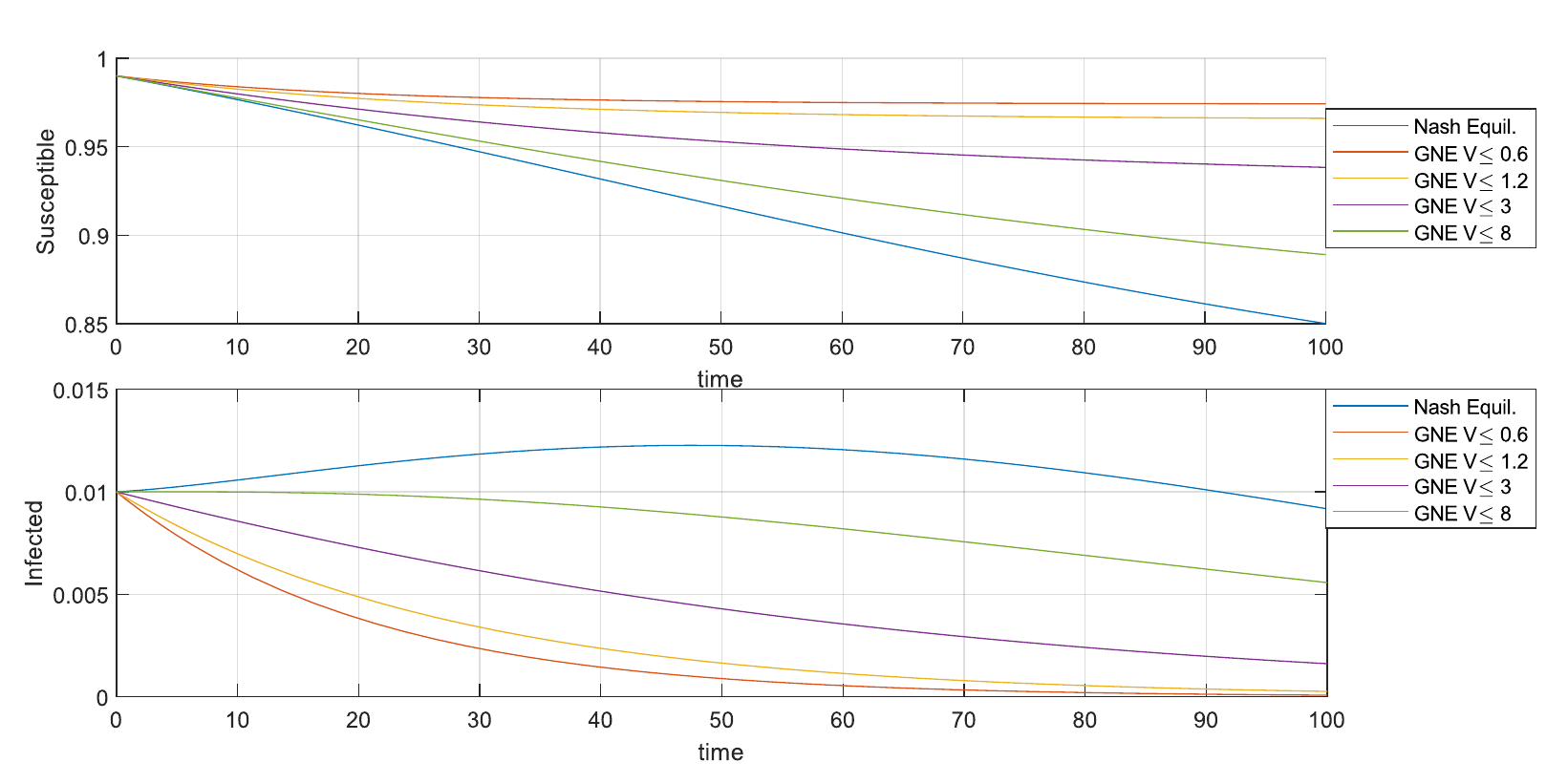}
		\caption{ The evolution over time for the total number of susceptible and infected persons for the Nash equilibrium and the GNE for various values of $C$. }
		\label{Epidemic_evol}
	\end{figure}

	The GNE under the constraint $V\leq C$, for various values of $C$ is illustrated in Figure \ref{many_GNEFig}. We observe that for some values of $C$ there are multiple GNE. 	Figure \ref{many_GNEFig1} shows how  costs of the non-vulnerable and vulnerable players vary as a function of $C$. We observe that, as the value of the constraint $C$ becomes smaller the cost of the vulnerable players decreases monotonically. Furthermore, compared to the unconstrained case, the cost of the non-vulnerable players under the variance constrained is improved as well.  Figure \ref{Epidemic_evol} illustrates  the evolution of the epidemic  for various values of $C$. We observe that, as the constraint becomes more restrictive i.e., as $C$ decreases the prevalence of the epidemic decreases as well. 

\section{Conclusion}

We analyzed social distancing games, involving vulnerable and non-vulnerable populations of players, characterized the Nash equilibria and investigated how inequality constraints influence the epidemic spread and the costs of the players.  
 We also defined a Generalized Nash equilibrium concept for non-atomic games with variance constraints, and characterized it in terms of single-point-supported deviations. Inequality constraints are always beneficial for the vulnerable players, and in some cases they could be beneficial for the non-vulnerable players as well. Furthermore, inequality constraints delay the spread of epidemics and 	reduce its prevalence. 

There are several directions for future research. First, the model can be generalized, including many classes of players having different vulnerabilities, minimum actions,  degrees, etc. Another direction is to  use real data to check the predictions of the model,  the modeling of the dynamic response of players and the study of other applications of the variance constrained games defined. 

\appendix 
\section{Proof of Proposition 1}
\label{PropExPr}

Equation  \eqref{meas_DE} can be written as:

\begin{equation}
\begin{aligned}
\dot x_{1u} &=-ru x_{1u}\cdot (\mathcal M x)\\
\dot x_{2u} &=-ru x_{2u}\cdot (\mathcal M x)\\
\dot x_{3u} &=ru x_{1u}\cdot (\mathcal M x) - \alpha_1x_{3u}\\
\dot x_{4u} &=ru x_{2u}\cdot (\mathcal M x)- \alpha_1x_{4u}\\
\dot x_{5} &=\mathcal M x
\end{aligned},\label{equat_Func_anal_fr}
\end{equation}
where $\mathcal M\in X^\star$ with:
$$\mathcal M x = \int  x_{3u} u \mu_1(du)+ \int  x_{4u} u \mu_2(du).$$
Note that $\|\mathcal M\|  \leq n_{1}+n_2$.
The initial conditions are $x_{1u}(0)=x_{2u}(0)=1-I_0$, $x_{3u}(0)=x_{4u}(0)=I_0$, for all $u\in[u_m,u_M]$ and $x_{5}(0)=0$.
In compact form we write $\dot x=f_{\mu_1,\mu_2}(x)$.

\begin{lemma}
	Any solution of \eqref{equat_Func_anal_fr} with the given initial conditions satisfies $0\leq x_{1u},\dots,x_{4u}\leq 1$. Let us denote this set by $X_0$ i.e., $X_0= \{x\in X: 0\leq x_{1u},\dots,x_{4u}\leq 1, \text{ for all } u\} $.
	\label{LEMMA_leq}
\end{lemma}
\textit{Proof}: Consider such a solution. Observe that, $x_{1u}(t),x_{2u}(t)\geq 0 $ for all $u$ and $t$.
Similarly, since $x\geq 0$ implies $\mathcal Mx\geq 0 $ we have  $x_{3u}(t),x_{4u}(t)\geq 0$  for all $t,u$. Finally, $\dot x_{1u}+\dot x_{3u}\leq 0$ and thus, $x_{3u}(t)\leq x_{1u}(0)+x_{3u}(0)=1$. Similarly, $x_{4u}(t)\leq  1$. \hfill $\square$

\begin{lemma}
	Let $x(t)$ be a  solution of   \eqref{equat_Func_anal_fr} with the given initial conditions. Then, $x_{ju}(t)$ is  continuous on $u$, for $j=1,\dots,4$.
	\label{LEMMA_cont}
\end{lemma}
\textit{Proof}: Let $h(t)=\mathcal M x(t)$. Then, $0\leq h(t)\leq n_1+n_2$. The solution of $\dot x_{1u} = -ruh(t)x_{1u}$ given by:
$$ x_{1u}(t)=(1-I_0)\exp\left( -  r u \int_0^t h(s)ds \right) $$
depends  continuously on $u$, $t$. A similar argument shows that $x_{2u}$ is continuous on $u$, $t$.  Then, observe that  the third equation of \eqref{equat_Func_anal_fr} can be written as:
$$\dot x_{3u} = b(u,t)- \alpha_1x_{3u},$$
where $b(u,t)=ru x_{1u} \cdot \mathcal M x $ is continuous on $u$ and bounded by $u_Mr(n_1+n_2)(1-I_0)$. The solution of this differential equation is:
$$ x_{3u}(t) = x_{3u}(0)e^{-\alpha_1t}+\int_{0}^te^{\alpha_1(\tau-t)}b(u,\tau)dt,
$$
which is again continuous on $u,t$. A similar argument shows that $x_{4u}$ is continuous on $u$, $t$.
\hfill $\square$

We then proceed to the proof of the proposition. Consider a saturated version of \eqref{equat_Func_anal_fr}:
\begin{equation}
\begin{aligned}
\dot x_{1u} &=-ru \text{sat}_1( x_{1u}) \cdot \text{sat}_2(\mathcal M x)\\
\dot x_{2u} &=-ru \text{sat}_1(x_{2u})\cdot \text{sat}_2(\mathcal M x)\\
\dot x_{3u} &=ru \text{sat}_1(x_{1u})\cdot \text{sat}_2(\mathcal M x) - \alpha_1x_{3u}\\
\dot x_{4u} &=ru \text{sat}_1(x_{2u})\cdot \text{sat}_2(\mathcal M x)- \alpha_2x_{4u}\\
\dot x_{5} &=\mathcal M x
\end{aligned},
\label{SaturatedSys}
\end{equation}
where $\text{sat}_1(z)=\max\{\min\{z,1\},0\}$ and $\text{sat}_2(z)=\max\{\min\{z,n_1+n_2\},0\}$. Due to Lemma \ref{LEMMA_leq}, any solution of \eqref{equat_Func_anal_fr}, with the given initial conditions,  is also a solution of the modified system.
Denote this system in compact form as $\dot x =\tilde f(x)$. It is not difficult to see that $\tilde f:X\rightarrow X$ is Lipschitz with constant: $$L  = 2ru_M(n_1+n_2)+\alpha_1+\alpha_2.$$ Thus, Theorem 7.3 of Brezis (2010)\cite{brezis2010functional}  implies that there exists a unique solution within the space $C([0,T],X)$ of continuous functions with values in $X$. Lemma \ref{LEMMA_cont}, along with the compactness of $[u_m,u_M]$ show  that there is no solution of \eqref{equat_Func_anal_fr}  not belonging to $C([0,T],X)$.\hfill $\square$

\begin{remark}
	Note that $\tilde f$ is uniformly Lipschitz. The constant is uniform in $\mu_1,\mu_2$, for positive measures of total mass $n_1,n_2$.
\end{remark}

\section{Proof of Lemma 1}

\label{Sect_3_lemma_Pr}

We then proceed to the proof Lemma \ref{Sect_3_lemma} and the computation of a formula for the directional derivative of the variance. To this end, we first consider the sensitivity of the solution of \eqref{meas_DE} with respect to the deviation $\varepsilon \delta  \mu_j$.

We first compute the directional derivative of $f_{\mu_1,\mu_2}(x)$ in the direction $\delta\mu$. The value of $f_{\mu_1,\mu_2}(x)$ depends on $\mu_j$ through the quantity  $I^f(t)$. This quantity can be written as:
$$I^f(t)=\mathcal {T}_{\mu_j} I_{j\cdot}(t)+ \mathcal {T}_{\mu_{-j}} I_{-j\cdot}(t),$$
where $ I_{j\cdot}(t):[u_m,u_M]\rightarrow\mathbb R$ with  $u\mapsto  I_{ju}(t)$
and
$\mathcal {T}_{\mu}\in(C([u_m,u_M],\mathbb R))^\star$  with:
\begin{equation}
\mathcal {T}_{\mu} I_{j\cdot}(t) = \int  I_{ju'}(t)u' \cdot\mu(du').\label{cal_T_def}
\end{equation}
The directional derivative of  $f_{\mu_1,\mu_2}(x)$ in the direction $\delta\mu_j$ is:
\begin{align*}
\mathcal D_{\delta \mu_j}& [ f_{\mu_1,\mu_2}(x)] =\\& [-ru  S_{ju}\mathcal {T}_{\delta \mu_j}  I_{j\cdot}, ~-ru  S_{(-j)u}\mathcal {T}_{\delta \mu_j}  I_{j\cdot}, ~ru  S_{ju}\mathcal {T}_{\delta \mu_j}  I_{j\cdot}, ~ru  S_{(-j)u}\mathcal {T}_{\delta \mu_j}  I_{j\cdot}, ~\mathcal {T}_{\delta \mu_j}  I_{j\cdot} ]^T.
\end{align*}
Observe that, due to the form of $\mathcal {T}_{\delta \mu_j}$ in \eqref{cal_T_def}, there is a bounded linear operator $\mathcal{B}_{\mu_1,\mu_2}(t):  Y\rightarrow X$ such that:
$$\mathcal D_{\delta \mu_j} [ f_{\mu_1,\mu_2}(x(t))] =\mathcal{B}_{\mu_1,\mu_2}(t) \delta \mu_j.$$

The linearized version of \eqref{meas_DE}, around the trajectories $( S_{1u},  S_{2u},  I_{1u},  I_{2u})$, is given by:
\begin{equation}
\begin{aligned}
\dot {x}'_{1u} &= -(ru I^f )x'_{1u}-ru  S_{1u}(\mathcal {T}_{\mu_1} x'_{3\cdot}+ \mathcal {T}_{\mu_{2}}x'_{4\cdot})-ru  S_{1u}\mathcal {T}_{\delta \mu_j}  I_{j\cdot} \\
\dot {x}'_{2u} &= -(ru I^f)x'_{2u}-ru  S_{2u}(\mathcal {T}_{\mu_1} x'_{3\cdot}+ \mathcal {T}_{\mu_{2}}x'_{4\cdot})-ru  S_{2u}\mathcal {T}_{\delta \mu_j}  I_{j\cdot} \\
\dot {x}'_{3u} &=  (ru I^f )x'_{1u}+ru  S_{1u}(\mathcal {T}_{\mu_1} x'_{3\cdot}+ \mathcal {T}_{\mu_{2}}x'_{4\cdot})-\alpha_1 {x'}_{3u}+ru  S_{1u}\mathcal {T}_{\delta \mu_j}  I_{j\cdot} \\
\dot {x}'_{4u} &= (ru I^f )x'_{2u}+ru  S_{2u}(\mathcal {T}_{\mu_1} x'_{3\cdot}+ \mathcal {T}_{\mu_{2}}x'_{4\cdot})-\alpha_{2}{x'}_{4u}+ru  S_{2u}\mathcal {T}_{\delta \mu_j}  I_{j\cdot}\\
\dot x'_5 &=  \mathcal {T}_{\mu_{1}}x'_{3u}+ \mathcal {T}_{\mu_{2}}x'_{4u}+\mathcal {T}_{\delta \mu_j}  I_{j\cdot} \label{linearized_variat}
\end{aligned},
\end{equation}
where $I^f=I^f_{\mu_1, \mu_2}$ as in \eqref{free_inf}. Thus, \eqref{linearized_variat} is a LTV system  in the form:
\begin{equation}
\dot  x' = \mathcal{A}_{\mu_1,\mu_2}(t) x' + \mathcal{B}_{\mu_1,\mu_2}(t) \delta \mu_j.
\label{LTV_SYS}
\end{equation}
\begin{lemma}
	\label{LinearizationLemma}
	Denote by $\phi_{\mu_1,\mu_2}(t)$ the solution of \eqref{meas_DE}. Then, the directional derivatives ${x}'_{lu}=\mathcal D_{\delta \mu_j} \phi^{lu}_{\mu_1,\mu_2}(t)$, for $l=1,\dots,4$ and ${x}'_{5}=\mathcal D_{\delta \mu_j} \phi^{5}_{\mu_1,\mu_2}(t)$ satisfy \eqref{LTV_SYS}, with zero initial conditions.
\end{lemma}
\textit{Proof:}
Consider the function:
$\bar f :Y\times X\rightarrow X$ with:
$$\bar f( \bar\mu ,x) =  f_{\bar\mu ,\mu_{-j}}(x)$$
and the set $D=X_0\times\{\mu\in Y:\|\mu\|= n_j\}$

The function $\bar f( \bar\mu ,x)$ is continuously Fr\'echet differentiable with respect to $x$. Its derivative is the operator $\mathcal{A}_{\mu_1,\mu_2}(t)$ given in
\eqref{linearized_variat}--\eqref{LTV_SYS}, which is continuous and bounded in $D$. The 	directional derivative: $$\mathcal D_{\delta \mu_j} [ \bar f( \bar\mu ,x) ] =\mathcal{B}_{\mu_1,\mu_2}(t) \delta \mu_j,$$
is also continuous in $x$ and bounded in $D$.
Furthermore, $X_0$ is positively invariant.
Thus,
Theorem 1 of
Banks et al. (2006)\cite{banks2006sensitivity} applies and the proof is complete.\hfill $\square$

Therefore, $F$ is continuous in $\mu_j$ and $J_j(u',\mu_j,\mu_{-j})$ is continuous in $\mu_j$, uniformly in $u'$.  This proves \eqref{Cost_variation}.

\begin{lemma}
	\label{IntFormLemma}
	There is a continuous function $ f^\text{var}_{j,F,\mu_1,\mu_2}(u)$
	such that the directional derivative of $F(\mu_1,\mu_2)$ on the $\delta \mu_j$ direction is given by:
	$$\mathcal D_{\delta \mu_j} F(\mu_1,\mu_2) = r x'_5(T)= \int f^\text{var}_{j,F,\mu_1,\mu_2}(u')\delta\mu_j(du').$$
\end{lemma}
\textit{Proof:}
Observe that $\mathcal{B}_{\mu_1,\mu_2}(t): Y\rightarrow X$ can be written as a    composition of two linear operators $\mathcal B^1_{\mu_1,\mu_2}(t):Y\rightarrow \mathbb R$ and $  \mathcal B^2_{\mu_1,\mu_2}(t):\mathbb R\rightarrow X$ with:
$\mathcal B_{\mu_1,\mu_2}^1(\delta\mu) = \mathcal {T}_{\delta \mu_j}  I_{j\cdot}$ and
$$\mathcal B_{\mu_1,\mu_2}^2(l) = \left [ \begin{matrix} -ru  x_{1u} l& -ru  x_{2u} l & ru  x_{1u} l&  ru  x_{2u} l &l\end{matrix}\right]^T.$$

The solution of the LTV system \eqref{LTV_SYS} with zero initial conditions  is written in the form (e.g. paragraph 1.4 of Deimling(2006)\cite{deimling2006ordinary}):
$$x'(T) = \int_0^T \Psi(T,t)\mathcal{B}_{\mu_1,\mu_2}(t)\delta\mu_j dt $$
where $\Psi(t,s):X\rightarrow X $ is the state transition operator.

Denote by $\mathcal C:X\rightarrow\mathbb R$ the operator  picking the last component i.e., $\mathcal C(x') = x'_5$. Then, $x'_5(T)$ can be written as:
$$x'_5(T)=\int_0^T [\mathcal C \Psi(T,t)\mathcal B^2_{\mu_1,\mu_2}(t)] \mathcal B^1_{\mu_1,\mu_2}(t)\delta\mu_j dt. $$
Note that  $\psi_t=\mathcal C \Psi(T,t) \mathcal B^2_{\mu_1,\mu_2}$ is a linear function $\psi_t:\mathbb R\rightarrow\mathbb R$. Write this function as $\psi_t(l)= \varphi (t) \cdot l$.
Thus:
$$x'_5(T) = \int_0^T \varphi(t) \int u x_{j+2,u}(t)\delta\mu_j(du) dt =  \int \left[u\int_0^T \varphi(t)  x_{j+2,u}(t) dt  \right]\delta\mu_j(du) $$
Define: $$f^\text{var}_{j,F,\mu_1,\mu_2}(u)=ru\int_0^T \varphi(t)  x_{j+2,u}(t) dt  $$
Observe that, since $x_{j+2,u}(t)$ is continuous with respect to $u$,  the function $f^\text{var}_{j,F,\mu_1,\mu_2}(u)$ is also continuous in $u$.
Thus:
$$\mathcal D_{\delta \mu_j} F(\mu_1,\mu_2) = rx'_5(T)= \int f^\text{var}_{j,F,\mu_1,\mu_2}(u)\delta\mu_j(du),$$ and the proof is complete \hfill $\square$

The directional derivative of $\bar u_j$ is given by:
$$\mathcal D_{\delta \mu_j} \bar u_j = \int  u' \delta \mu_j(du')/n_j.$$
The directional derivative  of the cost of  a player of type $j'$ who uses an action $u$ is given by:
\begin{align}
\mathcal D_{\delta \mu_j}  J_{j'} (u ,\mu_j , \mu_{-j}) &=  G_{j'}(1-I_0) ue^{-uF} \mathcal D_{\delta \mu_j} F-s_{j'j}u\mathcal D_{\delta \mu_j} \bar u_j\nonumber
\\&=  \int f^\text{var}_{j,J_{j'},\mu_1,\mu_2}(u',u)\delta\mu_j(du'),\label{D_Jj}
\end{align}
where:
$$
f^\text{var}_{j,J_{j'},\mu_1,\mu_2}(u',u) =  G_{j'}(1-I_0) ue^{-uF}f^\text{var}_{j,F,\mu_1,\mu_2}(u')-s_{j'j}uu'/n_j,
$$
is a continuous function of $u',u$.
The directional derivative of the mean cost of the players of type $j$ (the same with the type of the deviating players) is given by:
\begin{equation}
\begin{aligned}
\mathcal D_{\delta \mu_j} \bar J_j&=\lim_{\varepsilon\rightarrow0}\left[\frac{1}{\varepsilon n_j}\left[\int   J_j (u,\mu_j+ \varepsilon \delta  \mu_j,\mu_{-j}) (\mu_j+\varepsilon\delta\mu_j)(du)-    \int   J_j (u,\mu_j,\mu_{-j}) \mu_j (du)\right]\right]	\\
&= \int \mathcal  D_{\delta \mu_j} J_{j}  (u , \mu_j, \mu_{-j}) \bar \mu_j(du)+\lim_{\varepsilon\rightarrow0}\left[\frac{1}{ n_j}\int  J_j (u,\mu_j+ \varepsilon \delta  \mu_j,\mu_{-j}) \delta\mu_j (du)\right] \\
&= \int \mathcal D_{\delta \mu_j}  J_{j}  (u , \mu_j, \mu_{-j}) \bar \mu_j(du)+\frac{1}{n_j}\int  J_j (u,\mu_j,\mu_{-j}) \delta\mu_j (du) .
\end{aligned}\nonumber
\end{equation}
Substituting  \eqref{D_Jj} into the last equation and using Fubini's theorem, we get:
\begin{align}
\mathcal D_{\delta \mu_j} \bar J_j &= \int\left[ \frac{1}{n_j}   J_j (u,\mu_j,\mu_{-j})+   \int f^\text{var}_{J_{j},\mu_1,\mu_2}(u',u)          \bar \mu_j(du)  \right] \delta\mu_j(du')\nonumber \\
&=\int  f^\text{var}_{j,\bar J_{j},\mu_1,\mu_2}(u')  \delta\mu_j(du')
\end{align}
Similarly:
\begin{equation}
\begin{aligned}
\mathcal D_{\delta \mu_j} \bar J_{-j}= \int\left[   \int f^\text{var}_{j,J_{-j},\mu_1,\mu_2}(u',u)          \bar \mu_{-j}(du)  \right] \delta\mu_j(du')\nonumber =\int  f^\text{var}_{j,\bar J_{-j},\mu_1,\mu_2}(u')  \delta\mu_j(du')
\end{aligned}\nonumber
\end{equation}

Combining the last two equations we compute the variation of the mean cost:
$$\mathcal D_{\delta \mu_j} \bar J = (n_1\mathcal D_{\delta \mu_j} \bar J_1+n_2\mathcal D_{\delta \mu_j} \bar J_2)/(n_1+n_2)=\int  f^\text{var}_{j,\bar J,\mu_1,\mu_2}(u')  \delta\mu_j(du').  $$

We then proceed to the computation of the directional derivative of the variance in two steps. The first part is:
\begin{align*}
&\mathcal D_{\delta \mu_j} \left[\int  (J_j(u,\mu_j,\mu_{-j})-\bar J)^2\mu_j(du)\right] =\lim_{\varepsilon\rightarrow0} \frac{1}{\varepsilon} \left[\int  (J_j(u,\mu_j+\varepsilon\delta\mu_j,\mu_{-j})-\bar J(\mu_j+\varepsilon\delta\mu_j,\mu_{-j}))^2(\mu_j+\varepsilon\delta\mu_j)(du)-\right.
\\&~~~~~~~~~~~~~~~-\left.\int  (J_j(u,\mu_j,\mu_{-j})-\bar J(\mu_j,\mu_{-j}))^2\mu_j(du)\right]
\\&~~~= \int  (J_j(u,\mu_j,\mu_{-j})-\bar J)^2\delta\mu_j(du) +2\int   (J_j(u,\mu_j ,\mu_{-j})-\bar J(\mu_j,\mu_{-j}))    \int[ f^\text{var}_{j,J_{j},\mu_1,\mu_2}(u',u) -  f^\text{var}_{j,\bar J,\mu_1,\mu_2}(u') ] \delta\mu_j(du') \mu_j(du)
\\&~~~=\int    f^\text{var}_{j,V_{j},\mu_1,\mu_2}(u') \delta \mu_j(du').
\end{align*}
In the second equality the interchange of the integral with the limit is possible, because the convergence $\lim_{\varepsilon\rightarrow0}[J_j(u,\mu_j+\varepsilon\delta\mu_j,\mu_{-j})-J_j(u,\mu_j,\mu_{-j})]/\varepsilon$ is uniform in u. In the last equality we use again Fubini's theorem. Note that $f^\text{var}_{j,V_{j},\mu_1,\mu_2}(u')$ is continuous in $u'$.

Similarly:
\begin{align*}
&\mathcal D_{\delta \mu_j} \left[\int  (J_{-j}(u,\mu_j,\mu_{-j})-\bar J)^2\mu_j(du)\right]= \\&~~~~~~~~~=2\int   (J_{-j}(u,\mu_j ,\mu_{-j})-\bar J(\mu_j,\mu_{-j}))    \int[ f^\text{var}_{j,J_{-j},\mu_1,\mu_2}(u',u) -  f^\text{var}_{j,\bar J,\mu_1,\mu_2}(u') ] \delta\mu_j(du') \mu_j(du')
\\&~~~~~~~~~=\int    f^\text{var}_{j,V_{-j},\mu_1,\mu_2}(u') \delta\mu_j(du')
\end{align*}
Therefore, the directional derivative of the variance is written as:
$$\mathcal D_{\delta \mu_j} V(\mu_j,\mu_{-j}) =\int    f^\text{var}_{j,V,\mu_1,\mu_2}(u') \delta\mu_j(du'), $$
for a continuous function $f^\text{var}_{j,V,\mu_1,\mu_2}(u')$.

\section{Proof of Proposition 3}
\label{TwoPointPropPr}

Let us drop the dependence on $j$. 
Inequalities \eqref{Ineq1}, recalling that $\bar\mu'$ is a probability measure, can be written as:
$$ \int  \bar f_{\mathcal{K}}(u')  \bar\mu' (du') <0,$$
$$ \int  \bar f_{\mathcal{L}}(u')  \bar\mu' (du') < 0,$$
where $ \bar f_{\mathcal{K}}(u') =  J_j(u',\mu_1,\mu_2)-\int  J_j(u',\mu_1,\mu_2) \bar\mu (du')$,  $ \bar f_{\mathcal{L}}(u') =  f^\text{var}_{j,V_{j},\mu_1,\mu_2}(u')-\int  f^\text{var}_{j,V_{j},\mu_1,\mu_2}(u')  \bar\mu (du')$.

Using the (uniform) continuity of $\bar f_{\mathcal{K}}$, $\bar f_{\mathcal{K}}$ we have that for a set of $u_1,\dots,u_N$ and $c_1,\dots,c_N>0$ it holds:
$$  c_1\bar f_{\mathcal{K}}(u_1)+\dots +c_N\bar f_{\mathcal{K}}(u_N) <0,$$
$$ c_1\bar f_{\mathcal{L}}(u_1)+\dots +c_N\bar f_{\mathcal{L}}(u_N) <0,$$

If there is a point $u_k$ where $\bar f_{\mathcal{K}}(u_k)<0$ and $\bar f_{\mathcal{L}}(u_k)\leq 0$ or $\bar f_{\mathcal{K}}(u_k)\leq 0$ and $\bar f_{\mathcal{L}}(u_k)< 0$ we are done. Thus, assume that there is no such point. Excluding from the summation the terms where both $\bar f_{\mathcal{K}}(u_k)$ and $\bar f_{\mathcal{L}}(u_k)$ are non-negative the inequalities still hold true. Thus, assume that for any point $u_k$ either
$\bar f_{\mathcal{K}}(u_k)<0<\bar f_{\mathcal{L}}(u_k)$  or
$\bar f_{\mathcal{L}}(u_k)<0<\bar f_{\mathcal{K}}(u_k)$. Reordering the terms, the inequalities can be written as:
\begin{equation}
\begin{aligned}
x_1+\dots+x_n-y_1-\dots -y_m<0\\
-z_1-\dots-z_n+w_1+\dots +w_m<0
\end{aligned}
\end{equation}
where
\begin{align*}
x_i &= c_{k_i}\bar f_{\mathcal{K}}(u_{k_i}), \quad k_i \in \{k :c_k\bar f_{\mathcal{K}}(u_k)>0\}, \\
y_i &= |c_{k_i}\bar f_{\mathcal{K}}(u_{k_i})|, \quad k_i \in \{k :c_k\bar f_{\mathcal{K}}(u_k)<0\},\\
w_i &= c_{k_i}\bar f_{\mathcal{L}}(u_{k_i}), \quad k_i \in \{k :c_k\bar f_{\mathcal{L}}(u_k)>0\}, \\
z_i &= |c_{k_i}\bar f_{\mathcal{L}}(u_{k_i})|, \quad k_i \in \{k :c_k\bar f_{\mathcal{L}}(u_k)<0\}.
\end{align*}
Let $i_0$ be such that $y_{i_0}/w_{i_0}\geq y_i/w_i$ for all $i$ and denote $\alpha =y_{i_0}/w_{i_0}$.
\newline
\underline{Claim 1}: There is a $j_0$ such that $\alpha z_{j_0}>x_{j_0}$. Indeed if this in not true, and $\alpha z_{j }\leq x_{j}$ for all $j$, then multiplying the second relationship with $\alpha$ we get:
$$-\alpha z_1-\dots-\alpha z_n+\alpha w_1+\dots +\alpha w_m\geq   -x_1-\dots-x_n+y_1+\dots+y_m>0.$$
This completes the proof of the claim. \newline
\underline{Claim 2}: We may choose a $\rho \in[0,1]$ such that:
$$\rho x_{j_0}-(1-\rho)y_{i_0}<0$$
$$-\rho z_{j_0}+(1-\rho)w_{i_0}<0$$
Indeed for $\rho$ in the interval:
$$\frac{y_{i_0}}{\alpha z_{j_0}+y_{i_0}}<\rho<\frac{y_{i_0}}{x_{j_0}+y_{i_0}},$$
both inequalities are valid. The interval is not empty, since $\alpha z_{j_0}>x_{j_0}$.

Choose such a $\rho$. Then, using the form of $x,y,z,w$, there are $k_1,k_2$ such that:
$$  \rho c_{k_1}\bar f_{\mathcal{K}}(u_{k_1})+ (1-\rho) c_{k_2}\bar f_{\mathcal{K}}(u_{k_2}) <0,$$
$$  \rho c_{k_1}\bar f_{\mathcal{L}}(u_{k_1})+ (1-\rho) c_{k_2}\bar f_{\mathcal{K}}(u_{k_2}) <0,$$

Thus, taking $$\bar\mu'' = \frac{\rho c_{k_1}}{\rho c_{k_1}+(1-\rho) c_{k_2}}\mathcal{d} _{u_{k_1}} + \frac{(1-\rho) c_{k_2}}{\rho c_{k_1}+(1-\rho) c_{k_2}}\mathcal{d}_{u_{k_2}}, $$
where $\mathcal{d}_u$ is the Dirac measure at $u$, and the proof is complete.

\section{Computation of GNE}
\label{Appendix.GNEcomp}

Consider a discrete set $ U_d\subset [u_{m},u_M]$ and a grid $U_d\times U_d$ of pairs $(u_1,u_2)$. 
We first find the set of pairs $(u_1,u_2)$ on the grid, such that $V(n_1\mathcal d_{u_1},n_2\mathcal d_{u_2})=C$ approximately holds. The dynamics becomes:
\begin{equation}
\begin{aligned}
\dot S_{1u_1} &= -ru_1I^fS_{1u_1}, &
\dot S_{2u_2} &= -ru_2I^fS_{2u_2}, \\
\dot I_{1u_1} &=  ru_1I^fS_{1u_1}- \alpha_1 I_{1u_1},~~ &
\dot I_{2u_2} &=  ru_2I^fS_{2u_2}- \alpha_2 I_{2u_2} \\
\dot z&=I^f,
\end{aligned} \label{SIR_diff_eq_dirac},
\end{equation}
where $I^f(t) = u_1n_1I_{1u_1}(t)+u_2n_2I_{2u_2}(t).$

For every point on the grid that  $V(n_1\mathcal d_{u_1},n_2\mathcal d_{u_2})=C$ approximately holds, we  compute the function
$g^{\mathcal K}_{j,\mu_1,\mu_2}(u)$, for $u\in U$, as:
$$g^{\mathcal K}_{j,\mu_1,\mu_2}(u)=J_j(u,\mathcal d_{u_1},\mathcal d_{u_2})-J_j(u_j,\mathcal d_{u_1},\mathcal d_{u_2}).$$
For $j=1,2$ and a grid of values of $u\in U$, we solve the pair of differential equations:
\begin{align*}
\dot { S}_{1u} &= -ru I^f S_{1u}, &~
\dot { I}_{1u} &= ru I^f S_{1u} -\alpha_j{ I}_{1u}
\end{align*}
where $I^f$ is the quantity computed during the solution of \eqref{SIR_diff_eq_dirac}.

Consider the variation $\delta \mu_j = \mathcal d_u-\mathcal d_{u_j}$
The linearized system around the solution of \eqref{SIR_diff_eq_dirac} consists of $5$ differential equations:
\begin{equation}
\begin{aligned}
\dot {x}'_{1u_1} &= -(ru_1 I^f )x'_{1u_1}-ru_1  S_{1u_1}(n_1u_1 x'_{3u_1}+n_2u_2 x'_{4u_2})-ru_1   S_{1u_1}(   I_{ju}-   I_{ju_j}) \\
\dot {x}'_{2u_2} &= -(ru I^f)x'_{2u_2}-ru_2   S_{2u_2}(n_1u_1 x'_{3u_1}+n_2u_2 x'_{4u_2})-ru_2   S_{2u_2} (   I_{ju}-   I_{ju_j})\\
\dot {x}'_{3u_1} &=  (ru_1 I^f )x'_{1u_1}+ru_1  S_{1u_1}(n_1u_1 x'_{3u_1}+n_2u_2 x'_{4u_2})-\alpha_1 {x'}_{3u_1}+ru_1  S_{1u_1}  (  I_{ju}-  I_{ju_j})     \\
\dot {x}'_{4u_2} &= (ru_2 I^f )x'_{2u_2}+ru_2  S_{2u_2}(n_1u_1 x'_{3u_1}+n_2u_2 x'_{4u_2})-\alpha_{2}{x'}_{4u_2}+ru_2  S_{2u_2}(  I_{ju}-  I_{ju_j})      \\
\dot x'_5 &=  n_1u_1 x'_{3u_1}+n_2u_2 x'_{4u_2}+   I_{ju}-  I_{ju_j}
\end{aligned},\nonumber
\end{equation}
with zero initial conditions.
It holds $\mathcal D_{\delta \mu_j} F(\mu_1,\mu_2) = x'_5(T)$, and $\mathcal D_{\delta \mu_j} \bar u_j =   (u -u_j)/n_j.$ 
The directional derivative $\mathcal D_{\delta \mu_j}  J_{j'} (u ,\mu_j , \mu_{-j})$ is given in \eqref{D_Jj} and the directional derivative   $\mathcal D_{\delta \mu_j} \bar J_j$ by:
\begin{equation}
\begin{aligned}
\mathcal D_{\delta \mu_j} \bar J_j =  \mathcal D_{\delta \mu_j}  J_{j}  (u_j , \mu_j, \mu_{-j})  +\frac{1}{n_j} ( J_j (u,\mu_j,\mu_{-j}) - J_j (u_j,\mu_j,\mu_{-j}) ).
\end{aligned}\nonumber
\end{equation}
Furthermore  $
\mathcal D_{\delta \mu_j} \bar J_{-j} =  \mathcal D_{\delta \mu_j}  J_{-j}  (u_{-j} , \mu_j, \mu_{-j})$ and  $\mathcal D_{\delta \mu_j} \bar J = (n_1\mathcal D_{\delta \mu_j} \bar J_1+n_2\mathcal D_{\delta \mu_j} \bar J_2)/(n_1+n_2).  $
The directional derivative of the variance can be written as:  $$\mathcal D_{\delta \mu_j}V=(\mathcal D_{\delta \mu_j}V_1+\mathcal D_{\delta \mu_j}V_2)/(n_1+n_2),$$
where 
\begin{align*}
\mathcal D_{\delta \mu_j}V_1&=\mathcal D_{\delta \mu_j} \left[\int  (J_j(u',\mu_j,\mu_{-j})-\bar J)^2\mu_j(du')\right]  \\&= (J_j(u,\mu_j,\mu_{-j})-\bar J)^2-(J_j(u_j,\mu_j,\mu_{-j})-\bar J)^2+
\\& ~~~~~~+2 n_j (J_j(u_j,\mu_j ,\mu_{-j})-\bar J(\mu_j,\mu_{-j}))
\left[ \mathcal D_{\delta \mu_j}J_j(u_j,\mu_j,\mu_{-j})-\mathcal D_{\delta \mu_j}\bar J \right],
\end{align*}
and:
\begin{align*}
\mathcal D_{\delta \mu_j}V_2&=\mathcal D_{\delta \mu_j} \left[\int  (J_{-j}(u',\mu_j,\mu_{-j})-\bar J)^2\mu_{-j}(du')\right]  
\\& =2 n_{-j} (J_{-j}(u_{-j},\mu_j ,\mu_{-j})-\bar J(\mu_j,\mu_{-j}))
\left[ \mathcal D_{\delta \mu_j}J_{-j}(u_{-j},\mu_j,\mu_{-j})-\mathcal D_{\delta \mu_j}\bar J \right]
\end{align*}

We then compute the values of $g^{\mathcal L}_{j,\mu_1,\mu_2}(u)$, for $u\in U$  and use Corollary \ref{Coroll2} to check if $u_1,u_2$ is a GNE.

\nocite{*}
\bibliography{refs3.bib}%

\begin{thebibliography}{10}
\providecommand \doibase [0]{http://dx.doi.org/}%

\bibitem{ferguson2020report}
Ferguson N, Laydon D, Nedjati~Gilani G, et al. Report 9: Impact of
  non-pharmaceutical interventions (NPIs) to reduce COVID19 mortality and
  healthcare demand.  2020.

\bibitem{Kermack}
Kermack WO, McKendrick AG. A contribution to the mathematical theory of
  epidemics. {\it Proceedings of the royal society of london. Series A,
  Containing papers of a mathematical and physical character} 1927\string;
  115(772)\string: 700--721.

\bibitem{Ross}
Ross R. An application of the theory of probabilities to the study of a priori
  pathometry. {\it Proceedings of the Royal Society of London. Series A,
  Containing papers of a mathematical and physical character} 1916\string;
  92(638)\string: 204--230.

\bibitem{allen2008mathematical}
Allen LJ, Brauer F, Driessche V.~dP, Wu J. {\it Mathematical epidemiology}.
  1945.
\newblock Springer .
\newblock 2008.

\bibitem{Pastor-Satorras}
Pastor-Satorras R, Castellano C, Van~Mieghem P, Vespignani A. Epidemic
  processes in complex networks. {\it Reviews of modern physics} 2015\string;
  87(3)\string: 925.

\bibitem{Zhang1}
Zhang H, Zhang J, Zhou C, Small M, Wang B. Hub nodes inhibit the outbreak of
  epidemic under voluntary vaccination. {\it New Journal of Physics}
  2010\string; 12(2)\string: 023015.

\bibitem{Chang}
Chang SL, Piraveenan M, Prokopenko M. Impact of network assortativity on
  epidemic and vaccination behaviour. {\it arXiv preprint arXiv:2001.01852}
  2020.

\bibitem{Bauch1}
Bauch CT, Earn DJ. Vaccination and the theory of games. {\it Proceedings of the
  National Academy of Sciences} 2004\string; 101(36)\string: 13391--13394.

\bibitem{Bauch2}
Bauch CT, Galvani AP, Earn DJ. Group interest versus self-interest in smallpox
  vaccination policy. {\it Proceedings of the National Academy of Sciences}
  2003\string; 100(18)\string: 10564--10567.

\bibitem{Reluga1}
Reluga TC, Bauch CT, Galvani AP. Evolving public perceptions and stability in
  vaccine uptake. {\it Mathematical biosciences} 2006\string; 204(2)\string:
  185--198.

\bibitem{Reluga2}
Reluga TC, Galvani AP. A general approach for population games with application
  to vaccination. {\it Mathematical biosciences} 2011\string; 230(2)\string:
  67--78.

\bibitem{Zhang2}
Zhang H, Fu F, Zhang W, Wang B. Rational behavior is a ‘double-edged
  sword’when considering voluntary vaccination. {\it Physica A: Statistical
  Mechanics and its Applications} 2012\string; 391(20)\string: 4807--4815.

\bibitem{Fine-Clarkson}
Fine PEM, Clarkson JA. Individual versus public priorities in the determination
  of optimal vaccination policies. {\it American journal of epidemiology}
  1986\string; 124(6)\string: 1012--1020.

\bibitem{Kremer}
Kremer M. Integrating behavioral choice into epidemiological models of AIDS.
  {\it The Quarterly Journal of Economics} 1996\string; 111(2)\string:
  549--573.

\bibitem{Vardavas}
Vardavas R, Breban R, Blower S. Can influenza epidemics be prevented by
  voluntary vaccination?. {\it PLoS computational biology} 2007\string; 3(5).

\bibitem{Del_Valle}
Del~Valle S, Hethcote H, Hyman JM, Castillo-Chavez C. Effects of behavioral
  changes in a smallpox attack model. {\it Mathematical biosciences}
  2005\string; 195(2)\string: 228--251.

\bibitem{Chen2}
Chen FH. Rational behavioral response and the transmission of STDs. {\it
  Theoretical population biology} 2004\string; 66(4)\string: 307--316.

\bibitem{Funk-Review}
Funk S, Salath{\'e} M, Jansen VA. Modelling the influence of human behaviour on
  the spread of infectious diseases: a review. {\it Journal of the Royal
  Society Interface} 2010\string; 7(50)\string: 1247--1256.

\bibitem{Funk1}
Funk S, Gilad E, Watkins C, Jansen VA. The spread of awareness and its impact
  on epidemic outbreaks. {\it Proceedings of the National Academy of Sciences}
  2009\string; 106(16)\string: 6872--6877.

\bibitem{Chen1}
Chen FH. Modeling the effect of information quality on risk behavior change and
  the transmission of infectious diseases. {\it Mathematical biosciences}
  2009\string; 217(2)\string: 125--133.

\bibitem{d'Onofrio}
d’Onofrio A, Manfredi P. Information-related changes in contact patterns may
  trigger oscillations in the endemic prevalence of infectious diseases. {\it
  Journal of Theoretical Biology} 2009\string; 256(3)\string: 473--478.

\bibitem{theodorakopoulos2012selfish}
Theodorakopoulos G, Le~Boudec JY, Baras JS. Selfish response to epidemic
  propagation. {\it IEEE Transactions on Automatic Control} 2012\string;
  58(2)\string: 363--376.

\bibitem{trajanovski2015decentralized}
Trajanovski S, Hayel Y, Altman E, Wang H, Van~Mieghem P. Decentralized
  protection strategies against SIS epidemics in networks. {\it IEEE
  Transactions on Control of Network Systems} 2015\string; 2(4)\string:
  406--419.

\bibitem{hota2019game}
Hota AR, Sundaram S. Game-theoretic vaccination against networked SIS epidemics
  and impacts of human decision-making. {\it IEEE Transactions on Control of
  Network Systems} 2019\string; 6(4)\string: 1461--1472.

\bibitem{huang2019differential}
Huang Y, Zhu Q. A differential game approach to decentralized virus-resistant
  weight adaptation policy over complex networks. {\it IEEE Transactions on
  Control of Network Systems} 2019\string; 7(2)\string: 944--955.

\bibitem{toxvaerd2020equilibrium}
Toxvaerd F. Equilibrium social distancing. {\it Cambridge working papers in
  Economics} 2020.

\bibitem{karlsson2020decisions}
Karlsson CJ, Rowlett J. Decisions and disease: a mechanism for the evolution of
  cooperation. {\it Scientific Reports} 2020\string; 10(1)\string: 1--9.

\bibitem{amaral2020epidemiological}
Amaral MA, Oliveira dMM, Javarone MA. An epidemiological model with voluntary
  quarantine strategies governed by evolutionary game dynamics. {\it arXiv
  preprint arXiv:2008.05979} 2020.

\bibitem{ye2020modelling}
Ye M, Zino L, Rizzo A, Cao M. Modelling epidemic dynamics under collective
  decision making. {\it arXiv preprint arXiv:2008.01971} 2020.

\bibitem{kabir2020evolutionary}
Kabir KA, Tanimoto J. Evolutionary game theory modelling to represent the
  behavioural dynamics of economic shutdowns and shield immunity in the
  COVID-19 pandemic. {\it Royal Society open science} 2021\string; 7(9)\string:
  201095.

\bibitem{lagos2020games}
Lagos AR, Kordonis I, Papavassilopoulos G. Games of Social Distancing during an
  Epidemic: Local vs Statistical Information. {\it arXiv preprint
  arXiv:2007.05185} 2020.

\bibitem{van_Boven}
Boven vM, Klinkenberg D, Pen I, Weissing FJ, Heesterbeek H. Self-interest
  versus group-interest in antiviral control. {\it PLoS One} 2008\string; 3(2).

\bibitem{hardin1968tragedy}
Hardin G. The tragedy of the commons. {\it Science} 1968\string; 162\string:
  1243--1248.

\bibitem{alfaro2020social}
Alfaro L, Faia E, Lamersdorf N, Saidi F. Social interactions in pandemics:
  fear, altruism, and reciprocity. tech. rep., National Bureau of Economic
  Research;  2020.

\bibitem{Brown}
N.~Brown PN, Collins B, Hill C, Barboza G, Hines L. Individual Altruism Cannot
  Overcome Congestion Effects in a Global Pandemic Game. {\it arXiv preprint
  arXiv:2103.14538} 2020.

\bibitem{levin2020assessing}
Levin AT, Hanage WP, Owusu-Boaitey N, Cochran KB, Walsh SP, Meyerowitz-Katz G.
  Assessing the age specificity of infection fatality rates for COVID-19:
  systematic review, meta-analysis, and public policy implications. {\it
  European journal of epidemiology} 2020\string: 1--16.

\bibitem{hoffmann2021older}
Hoffmann C, Wolf E. Older age groups and country-specific case fatality rates
  of COVID-19 in Europe, USA and Canada. {\it Infection} 2021\string;
  49(1)\string: 111--116.

\bibitem{fehr1999theory}
Fehr E, Schmidt KM. A theory of fairness, competition, and cooperation. {\it
  The quarterly journal of economics} 1999\string; 114(3)\string: 817--868.

\bibitem{fowler2005egalitarian}
Fowler JH, Johnson T, Smirnov O. Egalitarian motive and altruistic punishment.
  {\it Nature} 2005\string; 433(7021)\string: E1--E1.

\bibitem{robson2017eliciting}
Robson M, Asaria M, Cookson R, Tsuchiya A, Ali S. Eliciting the level of health
  inequality aversion in England. {\it Health economics} 2017\string;
  26(10)\string: 1328--1334.

\bibitem{lunn2020motivating}
Lunn PD, Timmons S, Belton CA, Barjakov{\'a} M, Julienne H, Lavin C. Motivating
  social distancing during the Covid-19 pandemic: An online experiment. {\it
  Social Science \& Medicine} 2020\string; 265\string: 113478.

\bibitem{Reluga3}
Reluga TC. Game theory of social distancing in response to an epidemic. {\it
  PLoS computational biology} 2010\string; 6(5).

\bibitem{Poletti2}
Poletti P, Ajelli M, Merler S. Risk perception and effectiveness of
  uncoordinated behavioral responses in an emerging epidemic. {\it Mathematical
  Biosciences} 2012\string; 238(2)\string: 80--89.

\bibitem{Poletti3}
Poletti P, Caprile B, Ajelli M, Pugliese A, Merler S. Spontaneous behavioural
  changes in response to epidemics. {\it Journal of theoretical biology}
  2009\string; 260(1)\string: 31--40.

\bibitem{Kaznow}
Facchinei F, Kanzow C. Generalised Nash Equilibrium Problems. {\it Ann. Oper.
  Res.} 2010\string; 175\string: 177-211.

\bibitem{facchinei2007finite}
Facchinei F, Pang JS. {\it Finite-dimensional variational inequalities and
  complementarity problems}.
\newblock Springer Science \& Business Media .
\newblock 2007.

\bibitem{paccagnan2018nash}
Paccagnan D, Gentile B, Parise F, Kamgarpour M, Lygeros J. Nash and wardrop
  equilibria in aggregative games with coupling constraints. {\it IEEE
  Transactions on Automatic Control} 2018\string; 64(4)\string: 1373--1388.

\bibitem{jacquot2018nonsmooth}
Jacquot P, Wan C. Nonsmooth aggregative games with coupling constraints and
  infinitely many classes of players. {\it arXiv preprint arXiv:1806.06230}
  2018.

\bibitem{jacquot2019nonatomic}
Jacquot P, Wan C. Nonatomic Aggregative Games with Infinitely Many Types. {\it
  arXiv preprint arXiv:1906.01986} 2019.

\bibitem{singh2018linear}
Singh R, Wiszniewska-Matyszkiel A, others . Linear quadratic game of
  exploitation of common renewable resources with inherent constraints. {\it
  Topological Methods in Nonlinear Analysis} 2018\string; 51(1)\string: 23--54.

\bibitem{mas1984theorem}
Mas-Colell A. On a theorem of {Schmeidler}. {\it Journal of Mathematical
  Economics} 1984\string; 13(3)\string: 201--206.

\bibitem{shim2012influence}
Shim E, Chapman GB, Townsend JP, Galvani AP. The influence of altruism on
  influenza vaccination decisions. {\it Journal of The Royal Society Interface}
  2012\string; 9(74)\string: 2234--2243.

\bibitem{kordonis2020model}
Kordonis I. A Model for Partial Kantian Cooperation. In: Springer.  2020 (pp.
  317--346).

\bibitem{brezis2010functional}
Brezis H. {\it Functional analysis, Sobolev spaces and partial differential
  equations}.
\newblock Springer Science \& Business Media .
\newblock 2010.

\bibitem{banks2006sensitivity}
Banks HT, Nguyen HK. Sensitivity of dynamical systems to Banach space
  parameters. {\it Journal of mathematical analysis and applications}
  2006\string; 323(1)\string: 146--161.

\bibitem{deimling2006ordinary}
Deimling K. {\it Ordinary differential equations in Banach spaces}. 596.
\newblock Springer .
\newblock 2006.

\end{thebibliography}

\clearpage



\end{document}